\documentclass[authoryear,3p,twocolumn]{elsarticle}

\usepackage{natbib}

\bibpunct{(}{)}{;}{a}{,}{,}

\usepackage{graphicx}

\usepackage{amssymb,array}
\usepackage{amsmath,amsthm,wasysym}
\usepackage{setspace,lineno}

\begin{document}

\begin{frontmatter}



\title{Terrestrial Planet Evolution in the Stagnant-Lid Regime: \\
                        Size Effects and the Formation of Self-Destabilizing Crust}


\author{Joseph G. O'Rourke\footnote{Present address: Division of Geological and Planetary Sciences, California Institute of Technology, Pasadena, CA 91125, USA. jorourke@caltech.edu}}
\author{Jun Korenaga}

\address{Department of Geology and Geophysics, Yale University, New Haven, CT 06520, USA}

 \begin{abstract} \singlespacing
%
The ongoing discovery of terrestrial exoplanets accentuates the importance of studying planetary evolution for a wide range of initial conditions. We perform thermal evolution simulations for generic terrestrial planets with masses ranging from that of Mars to 10$M_\oplus$ in the stagnant-lid regime, the most natural mode of convection with strongly temperature-dependent viscosity. Given considerable uncertainty surrounding the dependency of mantle rheology on pressure, we choose to focus on the end-member case of pressure-independent potential viscosity, where viscosity does not change with depth along an adiabatic temperature gradient. We employ principal component analysis and linear regression to capture the first-order systematics of possible evolutionary scenarios from a large number of simulation runs. With increased planetary mass, crustal thickness and the degree of mantle processing are both predicted to decrease, and such size effects can also be derived with simple scaling analyses. The likelihood of plate tectonics is quantified using a mantle rheology that takes into account both ductile and brittle deformation mechanisms. Confirming earlier scaling analyses, the effects of lithosphere hydration dominate the effects of planetary mass. The possibility of basalt-eclogite phase transition in the planetary crust is found to increase with planetary mass, and we suggest that massive terrestrial planets may escape the stagnant-lid regime through the formation of a self-destabilizing dense eclogite layer.
\end{abstract}


\end{frontmatter}

\singlespacing
\section{Introduction}

Plate tectonics is only observed on Earth and is likely important to Earth's uniquely clement surface conditions \citep[e.g.,][]{Kasting2003}. Other terrestrial planets in the Solar System (i.e., Mercury, Mars, and Venus) are generally considered to feature a rigid spherical shell encompassing the entire planet, with hot mantle convecting beneath the shell \citep[e.g.,][]{STO2001}. This mode of mantle convection is known as stagnant-lid convection. In fact, stagnant-lid convection may be most natural for planetary mantles because the viscosity of constituent materials is strongly temperature-dependent \citep{Solo1995}. The discovery of many extrasolar terrestrial planets with mass 1~to 10$M_\oplus$ \citep[e.g.,][]{Rivera2005,Udry2007,Queloz2009,Mayor2009,Leger2009,Char2009,Borucki2011} makes understanding planetary evolution in the stagnant-lid regime especially critical.

Parametrized models of stagnant-lid convection have long been applied to planets in our Solar System in an effort to infer likely planetary evolution scenarios from limited observational constraints \citep[e.g.,][]{DJS83, Spohn91, HP2002, AFJK2010}. Previous studies of massive terrestrial planets are more theoretical in nature, focusing on two broad questions. First, the effects of planetary mass on the likelihood of plate tectonics have been studied through scaling analyses and simple parametrized convection models \citep{DV2007, COAL2007, JK2010, Heck2011}. Second, the evolution of planets in the stagnant-lid regime has been contrasted with evolution with plate tectonics in the hope of identifying atmospheric signatures that would indicate the regime of mantle convection for a distant planet \citep[e.g.,][]{Kite2009}. Mantle dynamics in the stagnant-lid regime, however, can be more complex than previously thought owing to the effects of mantle processing and crustal formation, and the scaling law of stagnant-lid convection that takes such complications into account has been developed only recently \citep{JK2009}. It is thus warranted to take a fresh look at the fate of massive terrestrial planets in the stagnant-lid regime and to explore the general effects of initial conditions including planetary mass.

This study extends a parametrized model of stagnant-lid convection recently applied to Mars \citep{AFJK2010} to terrestrial planets of various masses, including massive planets that evolve in the stagnant-lid regime that are termed ``super-Venus" planets. This model incorporates the effects of compositional buoyancy and dehydration stiffening on mantle dynamics \citep{JK2009}, which are rarely accounted for except in simulations of plate tectonics. Unlike in previous studies, sensitivity analyses are extensively performed to quantify the relationship between initial conditions and modeling results. Principal component analysis is used to simplify the interpretation of a large number of simulation results. Simple scaling analyses are also conducted to derive a theoretical basis for major modeling results. Moreover, the likelihood of plate tectonics is quantified by tracking the viscosity contrast across the lithosphere during each simulation.

The purpose of this study is to investigate paths along which generic terrestrial planets may evolve and to estimate whether massive terrestrial planets are relatively more or less likely to escape the stagnant-lid regime. Throughout this paper, ``Mars" and ``Venus" should be considered shorthand for generalized 0.107$M_\oplus$ and 0.815$M_\oplus$ terrestrial planets, respectively. The evolution of a particular planet is likely to diverge from the predictions of these simple parametrized models. Few constraints are available beyond planetary mass and radius for extrasolar terrestrial planets. But for terrestrial planets in our Solar System, more data are available from decades of observations and spacecraft visits. Here, we explore hypothetical planetary evolution with the simplest assumptions on mantle dynamics, thereby serving as a reference model on which additional complications may be considered if necessary.

\section{Theoretical Formulation}

Parametrized convection models are used to simulate the evolution of Mars, Venus, and putative super-Venus planets for a wide range of initial conditions. Equations used to track the thermal and chemical evolution of terrestrial planets are taken from \citet{AFJK2010} with some modifications. Earth-like, peridotite mantle compositions are used to parametrize melting behavior. Although continuous evolution in the stagnant-lid regime is assumed, a simple model of lithospheric weakening is also considered to evaluate the likelihood of plate tectonics occurring at some point during planetary evolution.

\subsection{Governing Equations}

Mars, Venus, and super-Venus planets are assumed to begin as differentiated bodies with a mantle and core. Energy conservation yields two governing equations. First, the energy balance for the core is \citep{DJS83}
\begin{equation} 
[4\pi R_i^2\rho_c(L_c+E_g)\frac{dR_i}{dT_{cm}}-\frac{4\pi}{3}R_c^3\rho_cC_c\eta_c]\frac{dT_{cm}}{dt} = 4\pi R_c^2F_c, 
\end{equation}
where $R_c$ and $R_i$ are the radii of the core and inner core, respectively; $\rho_c$ is the density of the core; $L_c$ is the latent heat of solidification associated with the inner core; $E_g$ is the gravitational energy liberated per unit mass of the inner core; $\eta_c$ is the ratio of $T_{cm}$, the temperature at the core side of the core/mantle boundary, to the average core temperature; $C_c$ is the specific heat of the core; and $F_c$ is the heat flux out of the core. The formulation of core cooling is identical to that of \citet{DJS83}.

Second, the energy balance for the mantle is \citep{HP2002}
\begin{eqnarray} 
\frac{4\pi}{3}(R_m^3 - R_c^3)\left(Q_m-\rho_mC_m\eta_m\frac{dT_u}{dt}\right)-\rho_mf_mL_m \nonumber \\
= 4\pi(R_m^2F_m-R_c^2F_c), 
\end{eqnarray}
where $R_m$ is the radius of the mantle; $Q_m$ is the volumetric heat production of the mantle; $\rho_m$ is the density of the mantle; $C_m$ is the specific heat of the mantle; $\eta_m$ is the ratio of the average temperature of the mantle to $T_u$, the potential temperature of the mantle (a hypothetical temperature of the mantle adiabatically brought up to the surface without melting); $f_m$ is volumetric melt production with associated latent heat release, $L_m$; and $F_m$ is the heat flux across the mantle/crust boundary. 

Some of the above parameters are universal constants, but most are planet-specific. Many important parameters are also time-varying. In particular, mantle melt is extracted to form crust, causing $R_m$ to decrease with time. Likewise, $Q_m$ decreases with time because of radioactive decay with some approximated average decay constant, $\lambda$ \citep{DJS83}, and extraction through mantle processing. 

\begin{figure*}[ht]
\centering
\includegraphics[width=6.25in]{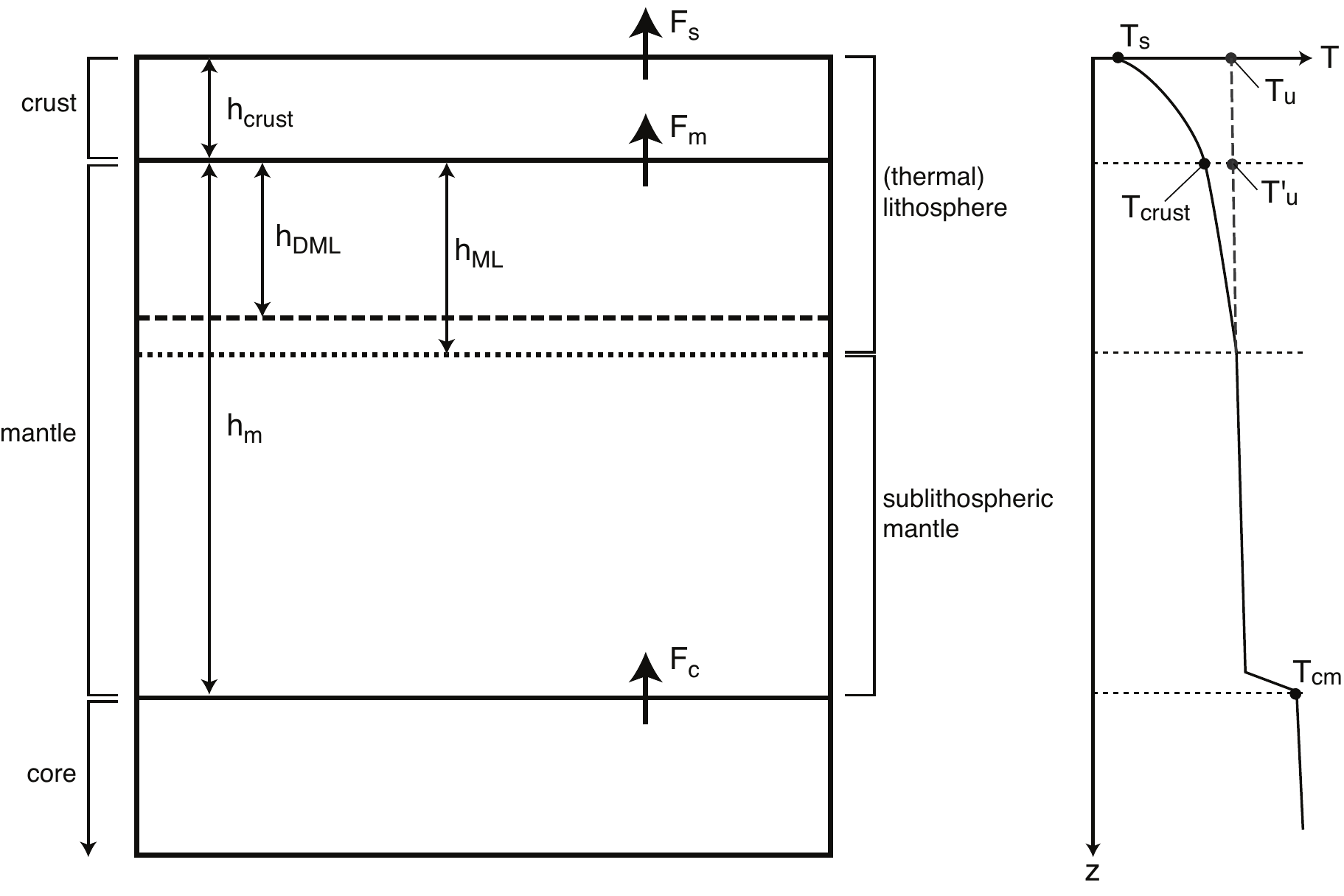}
\caption{Cartoons showing the assumed thermal and chemical structure of terrestrial planets taken from \citet{AFJK2010}. In general, terrestrial planets are divided into a crust, mantle, and core, as shown in the left panel. Mantle that has been processed by melting and stays in the thermal boundary layer is depleted mantle lithosphere (DML). The thickness of the DML must always be equal to or less than the thickness of the mantle lithosphere (ML). The section of the mantle below the thermal boundary layer is the sublithospheric mantle. The right panel shows the horizontally-averaged temperature distribution. Key model parameters are also indicated.} \label{fig:cartoon}
\end{figure*}

Figure~\ref{fig:cartoon} illustrates the assumed thermal and chemical structure in our model. Over time, melting processes an upper region of the original primitive mantle (PM) to form the crust and the depleted mantle lithosphere (DML). In parallel, the mantle lithosphere (ML), which is always thicker than the DML, develops as a conductive thermal boundary layer underlying the crust. As part of the DML can potentially delaminate and be mixed with the convecting mantle, the composition of the convecting mantle can be more depleted than that of the PM. The mantle below the DML is thus referred to as the source mantle (SM), the composition of which is initially identical to the composition of the PM but can deviate with time. The history of these layers strongly depends on convective vigor, effects of mantle melting, and initial conditions.

\subsection{Stagnant-Lid Convection with Mantle Melting}

Standard parameterizations are used for mantle rheology and the vigor of convection. Mantle viscosity is a function of mantle potential temperature and the degree of hydration as \citep{AFJK2010}
\begin{equation}
\eta(T_u,C_{SM}^W) = A\text{ exp}\left[\frac{E}{RT_u}+(1-C_{SM}^W)\text{log}\Delta \eta_w \right], \label{eq:visc}
\end{equation}
where $A$ is a constant factor calculated using a reference viscosity $\eta_0$ at a reference temperature $T^*_u$ = 1573~K; $E$ is the activation energy; $R$ is the universal gas constant; and $\Delta \eta_w$ is the viscosity contrast between wet and dry mantle. We use an activation energy of $E$ = 300 kJ mol$^{-1}$, which is appropriate for diffusion creep and dislocation creep within a Newtonian approximation \citep{Christensen1984,Karato1993,Korenaga2006}. The normalized water concentration in the source mantle, $C_{SM}^W$, has an initial value of one and decreases toward zero as mantle melting causes dehydration. To write Eq.~\ref{eq:visc}, we make the major assumption that mantle viscosity is not strongly pressure-dependent, which is consistent with early studies of the evolution of large rocky planets \citep[e.g.,][]{V06} and some theoretical predictions \citep{Karato2011} but in contrast to recent work \citep[e.g.,][]{Papuc2008,Vlada2011,Vlada2012}. Our choice is thus further explained in the discussion section.

Two non-dimensional parameters characterize thermal convection with the above viscosity formulation \citep{Solo1995}. First, the internal Rayleigh number serves to quantify potential convective vigor \citep{AFJK2010}
\begin{equation}
Ra_i = \frac{\alpha \rho_mg(T'_u-T_c)h_m^3}{\kappa \eta(T_u,C_{SM}^W)},
\end{equation}
where $\alpha$ is the coefficient of thermal expansion; $\kappa$ is the thermal diffusivity; $T_c$ and $T'_u$ are, respectively, the temperature at the bottom of the crust (called ``Moho temperature'') and the mantle potential temperature defined at the top of the mantle; and $h_m$ is the thickness of the mantle. Second, the Frank-Kamenetskii parameter is defined as \citep{Solo1995, AFJK2010}
\begin{equation}
\theta = \frac{E(T'_u-T_c)}{RT_u^2}.
\end{equation}

With these two parameters, the average convective velocity beneath the stagnant-lid may be calculated as \citep{SM2000}
\begin{equation}
u = 0.38\frac{\kappa}{h_m}\left(\frac{Ra}{\theta}\right)^{1/2}.
\end{equation}
To include the effects of compositional buoyancy and dehydration stiffening, the Nusselt number, which is a non-dimensional measure of convective heat flux, must be calculated with a local stability analysis at each time step \citep{JK2009}. The symbolic functionality may be expressed as
\begin{equation}
Nu = f(Ra,E,T_u,T_c,h_l,h_m,\Delta \rho,\Delta \eta_m),
\end{equation}
where $\Delta \rho$ and $\Delta \eta_m$ are the density and viscosity contrasts between the source mantle and depleted mantle, respectively, and $h_l$ is the thickness of the depleted mantle lithosphere. The thickness of a thermal boundary layer in the mantle is then easily calculated using
\begin{equation}
h_{ML} = \frac{h_m}{Nu}.
\end{equation}

The chemical evolution of the mantle strongly affects terrestrial planet evolution. To first order, partial melting of the mantle can be considered to begin at a depth where the temperature exceeds the solidus of dry peridotite, as long as the mantle is not significantly wet \citep{Hirth1996}. The initial pressure of melting is \citep{JK2002JGR}
\begin{equation}
P_i = \frac{T_u - 1423}{1.20\times10^{-7} - (dT/dP)_S},
\end{equation}
where $(dT/dP)_S$ is the adiabatic mantle gradient, which is roughly constant for the pressure range relevant to mantle melting. Therefore, $P_i$ should be approximately constant for any terrestrial planet with Earth-like mantle composition. Melting stops when the convective upwelling reaches the base of the mantle lithosphere. That is, the final pressure of melting is given by 
\begin{equation}
P_f = \rho_Lg(h_c+h_{ML}),
\end{equation}
where $h_c$ is the thicknesses of the crust; $g$ is gravitational acceleration; and $\rho_L$ is the density of the lithosphere. For convenience, we use the Martian $\rho_m$ as $\rho_L$, noting that $\rho_L$ should remain roughly constant whereas $\rho_m$, an averaged mantle parameter, increases with planetary mass because of pressure effects. If $P_f < P_i$, then melting occurs in the melting zone between $P_i$ and $P_f$, with thickness $d_m$ and average melt fraction equal to
\begin{equation}
\phi = \frac{P_i-P_f}{2}\left(\frac{d\phi}{dP}\right)_S,
\end{equation}
where $(d\phi/dP)_S$ is the melt productivity by adiabatic decompression. Volumetric melt productivity is finally parametrized as
\begin{equation}
f_m = \frac{2\chi d_mu\phi}{h_m}4\pi R_m^2, \label{eq:melt}
\end{equation}
where $\chi \sim 1$ if the upwelling mantle is cylindrical and all downwelling occurs at the cylinder's edge \citep{SM2000,AFJK2010}. The crustal temperature profile is calculated as in \citet{AFJK2010}, with the modification that crustal material with a temperature above $T_{crit}$ = 1273 K is considered to be buoyant melt that migrates within one time step immediately below the planet's surface, producing a relatively cooler crust and a larger mantle heat flux. This modification is not important for Martian cases, because the Moho temperature does not reach the threshold except for some extreme cases, but becomes essential to achieve a realistic crustal thermal profile for larger planets.

\subsection{Likelihood of Plate Tectonics}

Thermal evolution models featuring stagnant-lid convection are not applicable to planets on which plate tectonics occurs. If a suitable weakening mechanism exists, the lithosphere may be broken into plates and recycled into the mantle. Many aspects of plate tectonics on Earth, however, are not captured in current mathematical models \citep{Berco2003}. Quantifying the conditions under which plate tectonics is favored over stagnant-lid convection is likewise difficult, and the effect of planetary mass on the likelihood of plate tectonics has been controversial \citep{DV2007,JK2010,Heck2011}. Recent studies suggest, however, that the effects of planetary mass on yield and convective stresses may be dominated by uncertainties in other important planetary parameters, such as internal heating and lithosphere hydration \citep{JK2010,Heck2011}. 

This study uses a simple scaling that is consistent with current understanding of rock mechanics \citep{JK2010}, though the possibility of different lithosphere weakening mechanisms \citep[e.g.,][]{Berco2008} cannot be excluded. We assume that plate tectonics can occur if convective stress exceeds the brittle strength of lithosphere given by
\begin{equation}
\tau_y = c_0 + \mu \rho gz,
\end{equation}
where $c_0$ is the cohesive strength, $\mu$ is the effective friction coefficient, and $z$ is depth \citep{MS1998}. Experimental data indicate that the cohesive strength is negligible under lithospheric conditions, i.e., $c_0/(\mu \rho z) \ll 1$ \citep{Byer1978}. We use another non-dimensional parameter \citep{JK2010}:
\begin{equation}
\gamma = \frac{\mu}{\alpha(T_u-T_s)},
\end{equation}
where the relevant temperature difference is the difference between the mantle potential and surface temperatures. In the parameterized convection model formulated in the previous section, we separately consider the crust and the mantle, but when discussing the likelihood of plate tectonics using the scaling of \citet{JK2010JGR}, it is more convenient to treat the crust and mantle together, assuming that crustal rheology is similar to mantle rheology.

Detailed scaling analyses \citep{JK2010, JK2010JGR} show that the effective viscosity contrast across the lithosphere can be parameterized as
\begin{equation}
\Delta \eta_L = \text{exp}(0.327\gamma^{0.647}\theta_{tot}),
\end{equation}
where $\theta_{tot}$ is the Frank-Kamenetskii parameter defined using the total temperature difference explained above, i.e.,
\begin{equation}
\theta_{tot} = \frac{E(T_u-T_s)}{RT_u^2}.
\end{equation}
A transition from plate-tectonic to stagnant-lid convection can take place if the above viscosity contrast exceeds a critical value 
\begin{equation}
\Delta\eta_{L,crit} = 0.25Ra_{i,tot}^{1/2}, 
\end{equation}
where $Ra_{i,tot}$ is defined to incorporate surface temperature as
\begin{equation}
Ra_{i,tot} = \frac{\alpha \rho_mg(T_u-T_s)(h_c+h_m)^3}{\kappa \eta(T_u,C_{SM}^W)}.
\end{equation}

For each simulation, if $\Delta \eta_L/\Delta \eta_{L,crit}\le1$ at any time, then plate tectonics may have been favored at some point during the evolution of a given planet. The satisfaction of this criterion may strongly depend on the value of $\mu$, so a wide range of values should be tested. For silicate rocks, plausible values of $\mu$ range from 0.6 to 0.7, according to both laboratory studies \citep{Byer1978} and measurements of crustal strength on Earth \citep[e.g.,][]{Brudy1997}. Surface water, however, may lower these values substantially via thermal cracking and mantle hydration \citep{JK2007}.

\section{Numerical Models}

The parametrized model described above was used to calculate thermal histories of Mars- and Venus-like planets and 1 to 10$M_\oplus$ super-Venus planets, where the $\oplus$ subscript denotes parameters for Earth, for a duration of 4.5~Gyr using numerical integration with a time step of 1~Myr. A wide parameter space was explored by varying the initial mantle potential temperature, $T_u(0)$; the initial core/mantle boundary temperature, $T_{cm}(0)$; the initial volumetric heat production, $Q_0$; the reference mantle viscosity, $\eta_0$; and the viscosity contrast between dry and wet mantle, $\Delta\eta_w$. Previous work for Mars demonstrated that simulation results were not very sensitive to the degree of compositional buoyancy and other parameters \citep{AFJK2010}. Table~\ref{table:const_univ} lists model constants common to all simulations, and Table~\ref{table:const_planets} lists planet-specific ones. 

\begin{table}
\begin{center}
\begin{tabular}{*4{c}}
\hline
Constant & Value & Units & Ref. \\ \hline
$\lambda$ & $1.38\times10^{-17}$ & s$^{-1}$ & [1] \\
$k$ & 4.0 & W m$^{-1}$ K$^{-1}$ & [1] \\
$\alpha$ & $2\times10^{-6}$ & K$^{-1}$ & [1] \\
$\kappa$ & $10^{-6}$ & m$^2$ s$^{-1}$ & [1] \\
$\rho_L$ & 3527 & kg m$^{-3}$ & [1] \\
$L_m$ & $6.0\times10^5$ & J kg$^{-1}$ & [2] \\
$L_c+E_g$ & $1.0\times10^{-6}$ & J kg$^{-1}$ & [2] \\
$C_m$ & 1000$^a$ & J kg$^{-1}$ K$^{-1}$ & [3] \\
$C_c$ & 850$^a$ & J kg$^{-1}$ K$^{-1}$ & [3] \\
$\eta_m$ & 1.3$^a$ & N/A & [1] \\
$\eta_c$ & 1.2$^a$ & N/A & [1] \\
$(dT/dP)_S$ & $1.54\times10^{-8}$ & K Pa$^{-1}$ & [4] \\
$(d\phi/dP)_S$ & $1.20\times10^{-8}$ & Pa$^{-1}$ & [4] \\
\hline
\end{tabular}
\end{center}
\caption{Summary of universal constants used in all simulations. References: 1. \citet{DJS83}, 2. \citet{AFJK2010}, 3. \citet{NBS2011}, 4. \citet{JK2002JGR}. $^a$Mars has $C_m$ = 1149; $C_c$ = 571; $\eta_m$ = 1.0; and $\eta_c$ = 1.1 \citep{AFJK2010}. } \label{table:const_univ}
\end{table}

\begin{table*}[ht]
\begin{center}
\begin{tabular}{{c}*{9}{r}*{1}{c}}
\hline
Constant & Mars & Venus & $1M_\oplus$ & $2M_\oplus$ & $4M_\oplus$ & $5M_\oplus$ & $6M_\oplus$ & $8M_\oplus$ & $10M_\oplus$ & Units \\ \hline
$g$  & 3.70 & 8.87 & 10.0 & 13.6 & 18.6 & 20.7 & 22.6 & 26.0 & 29.1 & m s$^{-2}$ \\
$T_s$ & 220 & 730 & 300 & 300 & 300 & 300 & 300 & 300 & 300 &  K \\
$R_p$ & 3390 & 6050 & 6307 & 7669 & 9262 & 9821 & 10295 & 11072 & 11696 & km \\
$R_c$ & 1550 & 3110 & 3295 & 3964 & 4723 & 4986 & 5206 & 5564 & 5848 & km \\
$\rho_m$ & 3527 & 3551 & 4476 & 4951 & 5589 & 5845 & 6078 & 6497 & 6873 & kg m$^{-3}$ \\
$\rho_c$ & 7200 & 12500 & 12961 & 14882 & 17594 & 18698 & 19708 & 21530 & 23174 & kg m$^{-3}$ \\
$P_{cm}$ & 19 & 130 & 151 & 284 & 556 & 697 & 842 & 1144 & 1463 & GPa \\
$P_c$ & 40 & 290 & 428 & 821 & 1639 & 2067 & 2508 & 3431 & 4406 & GPa \\
\hline
\end{tabular}
\end{center}
\caption{Summary of planet-specific constants for Mars, Venus, and seven super-Venus planets. Martian values were taken from \citet{AFJK2010} and references therein. Venusian values can be found in  \citet{Spohn91} and \citet{NBS2011}. Super-Venus values were calculated in this study from simple interior models following \citet{SS007}.} \label{table:const_planets}
\end{table*}

\subsection{Application to Mars- and Venus-like Planets}

For Venus, the following sets of initial conditions were used: Initial mantle potential temperature, $T_u(0)$ = 1400, 1550, 1700, 1850, and 2000~K; initial core/mantle boundary temperature, $T_{cm}(0)$ = 3500, 4000, and 4500~K; reference viscosity, $\eta_0$ = 10$^{18}$, 10$^{19}$, and 10$^{20}$~Pa s; and dehydration stiffening, $\Delta\eta_w$ = 1, 10, and 100. For Mars, initial core/mantle boundary temperatures were 2250, 2500, and 3000~K. In addition, five different values were tested for the amount of internal heating $Q_0$. Compositional buoyancy was set as $(d\rho/d\phi)$ = 120 kg m$^{-3}$ for all simulations. The fraction of light elements in the core was fixed at 0.2 for all simulation runs to avoid inner core solidification \citep{Schubert1992, AFJK2010}. Simulations were performed for all permutations of the above initial conditions, although unrealistic simulation results were discarded in the following way: For both Venus and Mars, inner core growth was disallowed and total surface heat flux at the present was required to be positive. Furthermore, the condition $h_c(t_p) <$ 500 km was imposed to disregard results with unrealistic crustal growth. None of the 675 simulations for Venus failed these criteria, but 30 of the 675 simulations for Mars were discarded.

The appropriate magnitude of radiogenic heating is poorly constrained in general, especially for terrestrial exoplanets. Even for Earth, the abundance of radiogenic heating is controversial. Geochemical constraints support a low Urey ratio, the ratio of internal heat production to surface heat flux, but this is known to conflict with the cooling history of Earth unless a non-classical heat-flow scaling for mantle convection is assumed \citep{Christ1985,JK2008RG}. A Urey ratio close to one has thus long been preferred from a geophysical perspective \citep{Davies1980,GS1980,STO2001} and can be used to provide an upper bound for the initial concentration of radioactive elements in Earth's chemically undifferentiated mantle. Assuming a present-day surface heat flux of 46 TW \citep{Jaupart2007}, an extreme upper bound for Earth is $Q_0 \approx$ 3.5$\times10^{-7}$ W m$^{-3}$.

A recent petrological estimate on the thermal history of Earth is actually shown to favor a low Urey ratio ($\sim$0.3) with a non-classical heat-flow scaling \citep{Herz2010}, indicating that geochemical constraints on the heat budget may be robust. In the thermal evolution models of \citet{Kite2009}, for example, concentrations of $^{40}$K, $^{232}$Th, $^{235}$U, and $^{238}$U taken from \citet{Ring1991} and \citet{TS2002} were considered, corresponding to values for $Q_0$ between $1.2\times10^{-7}$ and $8.7\times10^{-8}$ W m$^{-3}$ for Venus. We thus chose to use the following values for initial volumetric radiogenic heating: $Q_0$ = 0.5, 0.75, 1.0, 1.25, and 1.75 $\times10^{-7}$ W m$^{-3}$ (in the case of Venus). The default intermediate value is 1.0 $\times10^{-7}$ W m$^{-3}$. For other planets, $Q_0$ was multiplied by $\rho_m$/$\rho_{m,\venus}$, where $\venus$ indicates the Venusian value, to maintain constant element abundances in more or less compressed mantles. 

\subsection{Application to Super-Venus Planets}

One-dimensional profiles of massive terrestrial exoplanets were generated to calculate planet-specific constants used in the above stagnant-lid convection model. Many interior structure models exist for massive solid exoplanets, ranging from simple to very complex \citep{V06,SS007,CS007,Wag2011}. To study the first-order effects of planetary mass on stagnant-lid convection, a relatively simple structure with an Fe($\epsilon$) core and a MgSiO$_{\textrm{3}}$ mantle is assumed, as in \citet{SS007} and \citet{Kite2009}. The resulting interior density and pressure distributions neglect several obvious factors such as temperature effects, but yield results remarkably similar to those from more complex models.

Three equations are solved to calculate $m(r)$, the mass contained within radius $r$; $P(r)$, the pressure distribution; and $\rho(r)$, the density distribution.  A self-consistent internal structure must satisfy the material specific equations of state
\begin{equation} P(r) = f_{EOS}(\rho(r),T(r)), \end{equation}
the equation of hydrostatic equilibrium
\begin{equation} \frac{dP(r)}{dr} = \frac{-Gm(r)\rho(r)}{r^{2}}, \label{eq:hydrostatic} \end{equation}
and the conservation of mass equation for a spherical mass distribution
\begin{equation} \frac{dm(r)}{dr} = 4\pi r^{2}\rho(r), \label{eq:mass} \end{equation}
where $G$ is the gravitational constant; $T(r)$ is the radial temperature profile; and $f_{EOS}$ represents a material-specific equation of state \citep{SS007}.

\begin{table*}[ht]
  \begin{center}
 \begin{tabular}{*{6}{l}} \hline
   Material & $K_{0}$ (GPa) & $K^{'}_{0}$ & $K^{''}_{0}$ (GPa$^\textrm{-1}$) & $\rho_{0}$ (kg m$^\textrm{-3}$) & EOS \\ \hline
   Fe($\epsilon$) & 156.2 & 6.08 & N/A & 8300 & V \\
   MgSiO$_{\textrm{3}}$(pv) & 247 & 3.97 & -0.016 & 4100 & BME4 \\
   MgSiO$_{\textrm{3}}$(en) & 125 & 5 & N/A & 3220 & BME3 \\
   \hline
  \end{tabular}
  \end{center}
 \caption{Material constants used to generate interior structure models, taken from \citet{SS007}. Using three different equations of state, $P(\rho)$ is calculated to high resolution for each material. The Vinet and 3rd and 4th order Birch-Murnaghan equations of state are abbreviated V, BME3, and BME4, respectively.} \label{table:EOS}
\end{table*}

Equations of state are numerically calculated using constants from Table~\ref{table:EOS} to sufficient resolution so that $\rho(r)$ can be determined to within $\pm$1 kg $\textrm{m}^{-3}$. For a desired $M_P$, equations (\ref{eq:hydrostatic}) and (\ref{eq:mass}) are numerically integrated from the center of a planet with the inner boundary conditions $M(0) = 0$ and $P(0) = P_{c}$, where $P_{c}$ is a guessed central pressure. The outer boundary condition is simply $P(R_{P}) = 0$. Errors associated with ignoring temperature effects are limited to a few percent \citep{SS007}. With this method, the choice of $P_{c}$ determines the $R_P$ at which the outer boundary condition is satisfied. These calculations are iterated with the bisection method until $P_{c}$ is found such that $m(R_P) = M_P$ to within 0.1\%. The equation of state for Fe($\epsilon$) is used until $m(r) = 0.325M_P$, mandating a 32.5\% core mass fraction. The MgSiO$_{\textrm{3}}$ perovskite to enstatite phase transition is assumed to occur at 23 GPa, although the transition pressure increases to $\sim$25 GPa at $\sim$800 K \citep[e.g.,][]{Ita1992}. Neglecting this phase transition would produce unrealistically high near-surface densities. 

\begin{figure}
\centering
\includegraphics[width=20pc]{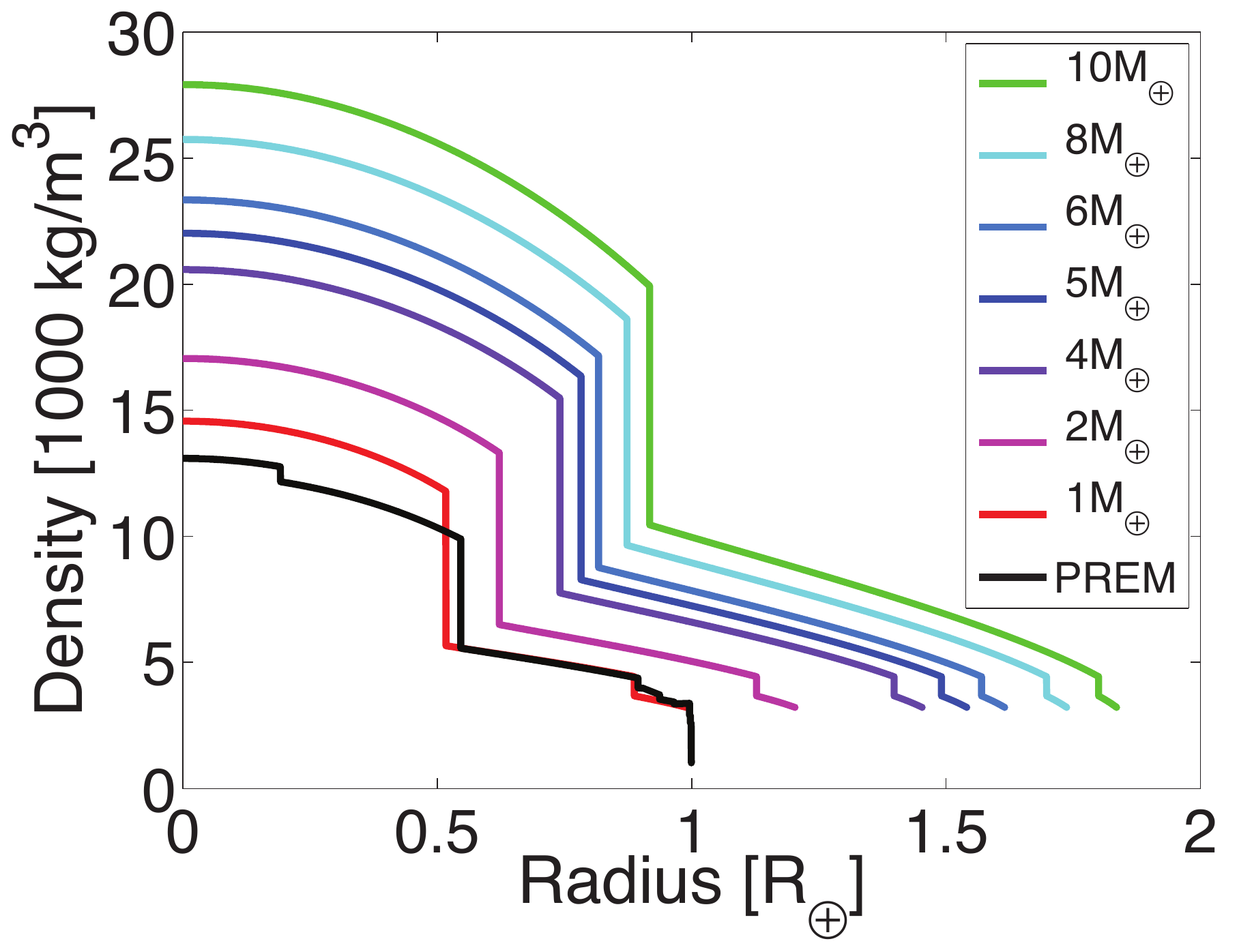}
\caption{Interior density distributions for super-Venus planets with $M_P =$ 1, 2, 4, 5, 6, 8, and 10$M_\oplus$.  The material equations of state and the equations of conservation of mass and hydrostatic equilibrium were numerically integrated to build simple planets, and the interior boundary condition was adjusted until the resulting planet had the desired mass. These simple models are used to calculate averaged values for mantle density, $\rho_m$; core density, $\rho_c$; surface gravity, $g$; central pressure, $P_c$; and core/mantle boundary pressure, $P_{cm}$. For comparison, the density distribution for Earth from the Preliminary Reference Earth Model (PREM) of \citet{PREM} is also plotted.} \label{fig:profile}
\end{figure}

Pressure, mass, and density distributions were calculated for planets with $M_P =$ 1, 2, 4, 5, 6, 8, and 10$M_\oplus$. From these models, averaged densities for the core and mantle were calculated, along with $P_c$ and $P_{cm}$. Surface gravitational accelerations were calculated using $g = GM_{P}/R_P^2$ for each planet. These constants are reported in Table~\ref{table:const_planets}. The density distributions for these planets are shown in Fig.~\ref{fig:profile}, along with the density profile for Earth from the Preliminary Reference Earth Model (PREM) of \citet{PREM}. Compared to PREM, the scheme for calculating internal structures used in this study overestimates the density of the core and underestimates the radius of the core of an Earth-mass planet, but returns $R_P(M_\oplus) \approx R_\oplus$ despite ignoring details of mineral composition, phase transitions, and temperature effects. In more massive planets, the enstatite to perovskite phase transition occurs at much shallower depths because a higher surface gravity causes a greater increase in pressure with depth. Furthermore, $P_c$ increases much more rapidly than $P_{cm}$ with increasing planetary mass.

Most of the Martian and Venusian initial conditions can be used for super-Venus thermal evolution models, but some must be modified appropriately. For example, the core/mantle temperature for super-Venus planets should increase along the mantle adiabatic temperature gradient: for 5 and 10$M_\oplus$ super-Venus planets, initial core/mantle boundary temperatures are increased by roughly 350 and 900 K, respectively, from the initial conditions for Venus. These temperatures still correspond to a so-called ``hot start," which is likely for terrestrial planets because of the large magnitude of gravitational potential energy released during accretion \citep{DJS83}. Only three simulations for the 5$M_\oplus$ planet, and no simulations for the 10$M_\oplus$ planet, failed the requirements on crustal thickness, surface heat flux, and inner core growth.

\section{Results}

Thermal evolution simulations were performed for Mars- and Venus-like planets and super-Venus planets. The following sections summarize the results, beginning with a few representative examples for Mars and Venus. Principal component analysis, as described in the appendix, was applied using all simulation results to identify major model behaviors. We also tried to quantify relations between input and output parameters and, despite the complexity of our model formulation, a linear function of initial conditions is found to reasonably approximate many output parameters of interest.

\subsection{Sample Thermal Histories for Mars- and Venus-like Planets}

\begin{figure*}[ht]
\centering
\includegraphics[width=39pc]{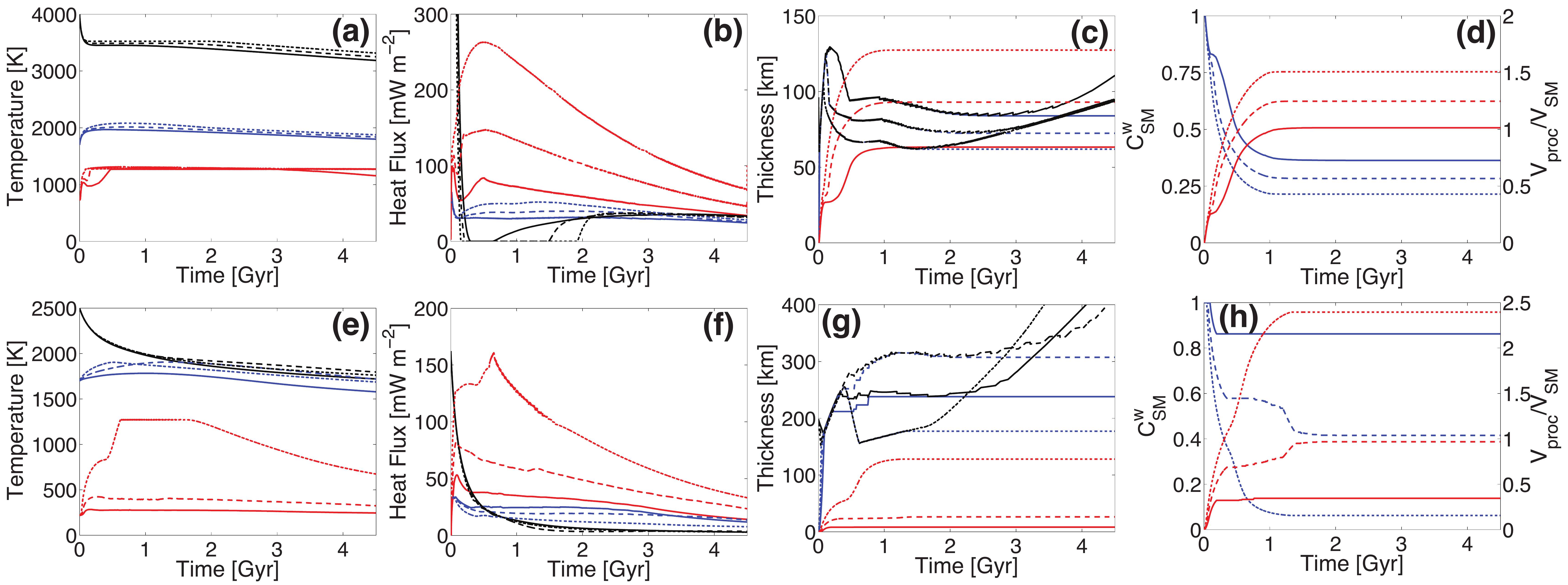}
\caption{Sample histories for Venus (top) and Mars (bottom). From left to right, red curves signify Moho temperature, surface heat flow, crustal thickness, and normalized mantle water content. Likewise, blue curves represent mantle potential temperature, mantle heat flux, depleted mantle lithosphere thickness, and fraction of processed source mantle; green curves show core/mantle boundary temperature, core heat flux, and lithosphere thickness. Solid, dashed, and dotted lines indicate Venusian $Q_0$ = 0.5, 1.0, and 1.75 $\times10^{-7}$ W m$^{-3}$, respectively. Default initial conditions are $T_u(0) = 1700$ K, $\eta_0 = 10^{19}$ $\text{Pa}\cdot\text{s}$, $\Delta\eta_w=100$, and $(d\rho/d\phi) = 120$ kg m$^{-3}$. Venus and Mars have $T_{cm}(0) =$ 4000 and 2500 K, respectively. Because crustal melting causes highly discontinuous surface heat flux, a moving average with a 75 Myr span was used for plotting purposes.} \label{fig:V&M_hist}
\end{figure*}

Sample thermal histories for Mars and Venus are shown in Fig.~\ref{fig:V&M_hist}. These models span the entire range of initial radiogenic heating values with all other initial conditions set to intermediate values. In particular, $T_u(0)$ = 1700 K, $\mu_0$ = 10$^{19}$ Pa s, and $\Delta\mu_w$ = 100. For Venus and Mars, respectively, $T_{cm}(0)$ = 4000 and 2500 K and $T_s$ = 220 and 730 K. Initially very hot cores are assumed here because core segregation is expected to release a large amount of gravitational potential energy. This excess heat is released into the mantle during the first hundred million years of planetary evolution. Thereafter, mantle dynamics controls core cooling. Whereas internal heating has a great effect on surface heat flux, mantle temperatures only differ to within $\pm$200 K for the sampled range of $Q_0$. Mars evolves with a consistently lower potential temperature than Venus. Because Mars also has a relatively shallow mantle, the Martian core is cooled down more efficiently.

Crustal thickness is an important, potentially observable constraint for planetary evolution models. Mars and Venus, with different magnitudes of radiogenic heating, have very different crustal formation histories. Both planets start with no initial crust, but quickly produce some through mantle melting. For Venus, Moho temperatures quickly reach the melting point of basalt for all initial internal heating choices. Crustal production occurs for the first $\sim$1 Gyr of evolution, with thicker crust for higher internal heating. For Mars, crustal production is gradual and crustal temperatures are much lower, with increased internal heating causing an longer period of crustal formation and increased total crustal production. Both Mars and Venus undergo substantial mantle processing, indicating that the deep interior serves as a significant source of endogenous water, especially during the first $\sim$1.5 Gyr of their evolution.

\subsection{Sensitivity Analyses for the Evolution of Mars- and Venus-like Planets}

\begin{figure*}[ht]
\centering
\includegraphics[width=39pc]{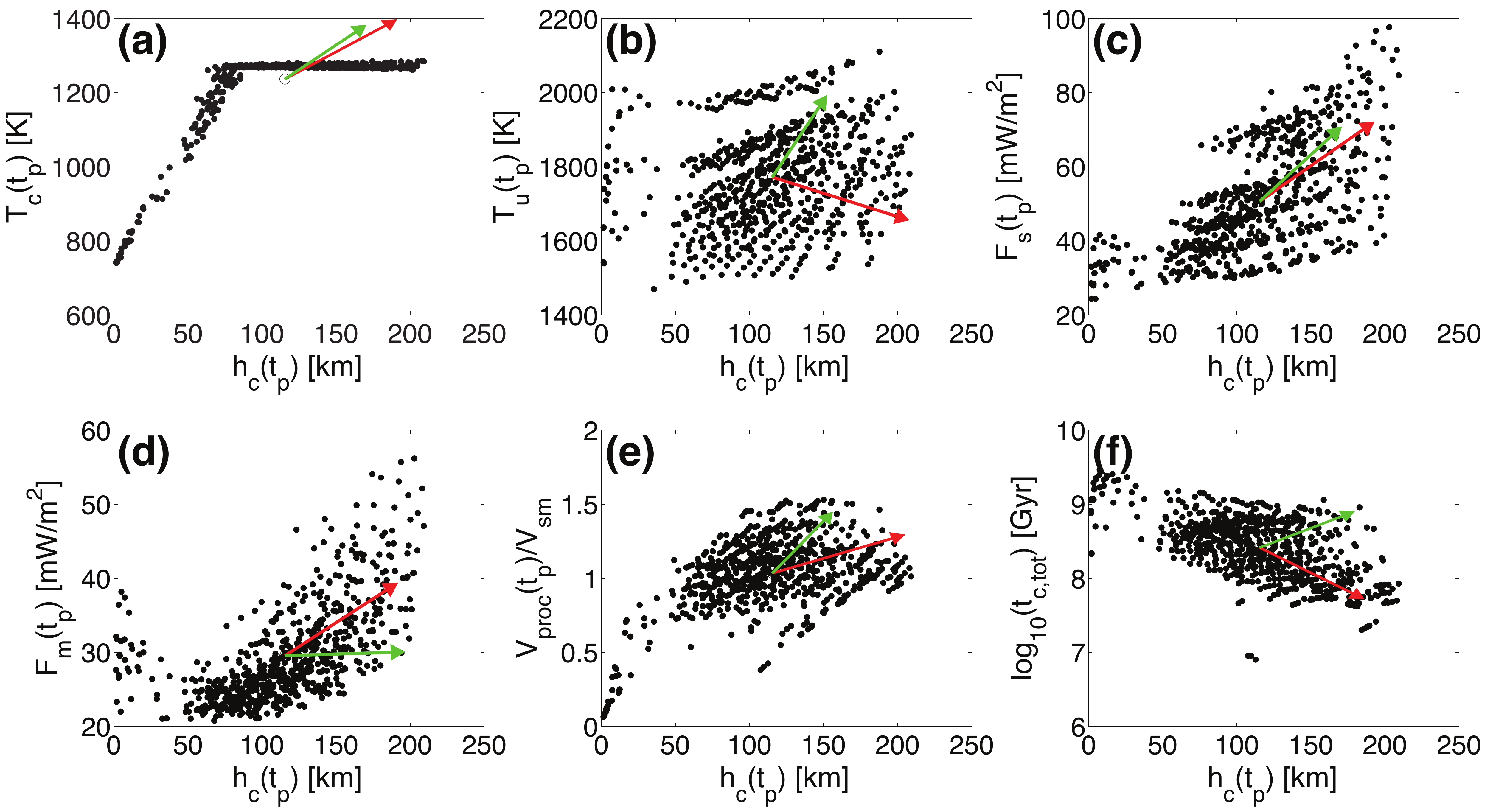}
\caption{Summary of parameter values at the present for 675 simulations of the thermal evolution of Venus. Arrows are projections of the principal component basis vectors that emanate from a point representing the averaged simulation results, indicating axes that account for the vast majority of the data set's variance. The red arrow represents a larger percentage of cumulative variance (41\%) than the green arrow (27\%). Panels show (a) Moho temperature, (b) mantle potential temperature, (c) surface heat flux, (d) mantle heat flux, (e) fraction of mantle processed by melting, and (f) total time for crust to grow from 10\% to 95\% of its present thickness as functions of crustal thickness. Because crustal melting causes highly discontinuous surface and mantle heat fluxes, the model outputs are the averaged values for the final 100 Myr of planetary evolution.} \label{fig:Venus_PCA}
\end{figure*}

\begin{figure*}[ht]
\centering
\includegraphics[width=39pc]{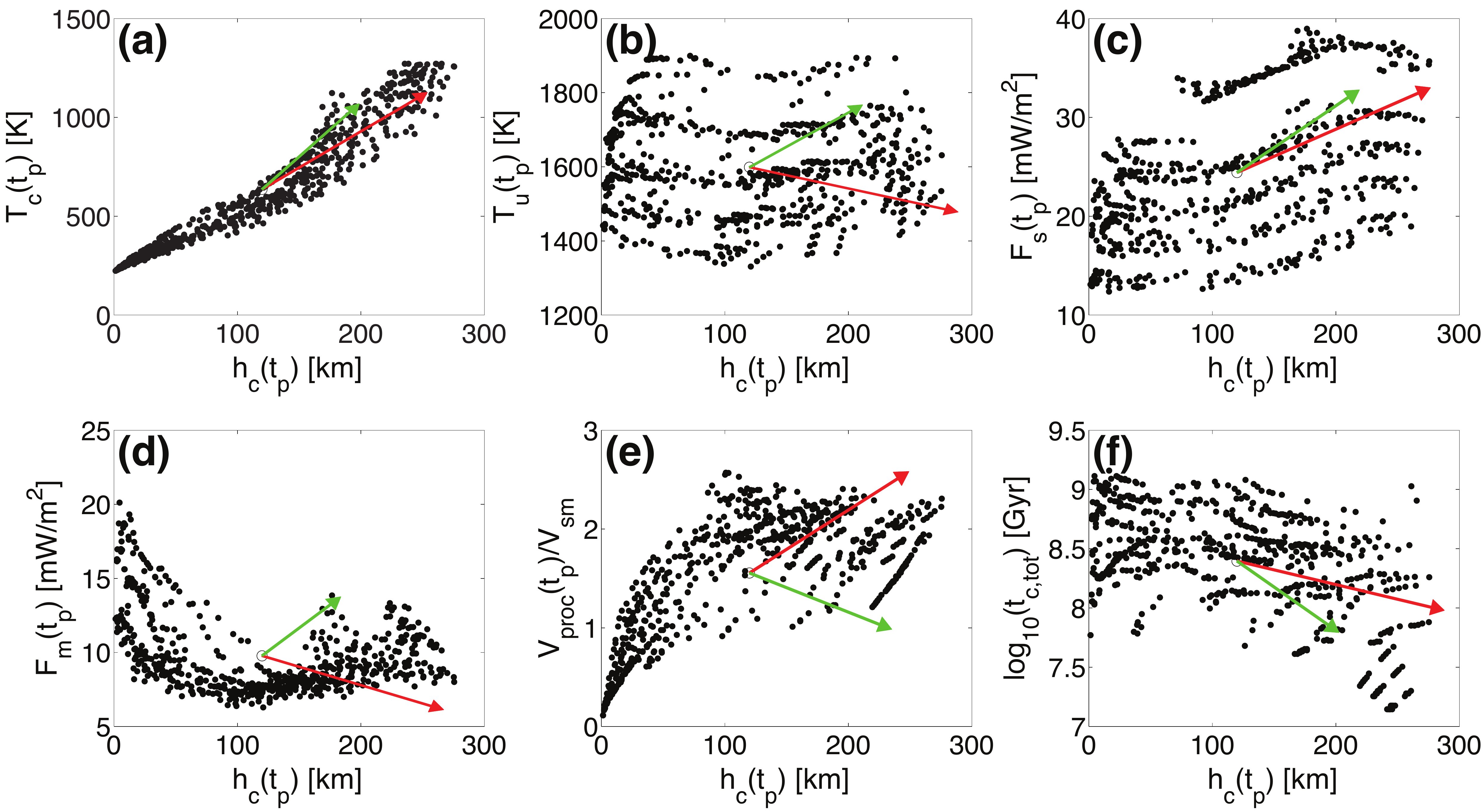}
\caption{Summary of parameter values at the present for 645 simulations of the thermal evolution of Mars. Arrows are projections of the principal component basis vectors that emanate from a point representing the averaged simulation results, indicating axes that account for the vast majority of the data set's variance. The red arrow represents a larger percentage of cumulative variance (42\%) than the green arrow (23\%). Panels show (a) Moho temperature, (b) mantle potential temperature, (c) surface heat flux, (d) mantle heat flux, (e) fraction of mantle processed by melting, and (f) total time for crust to grow from 10\% to 95\% of its present thickness as functions of crustal thickness. Because crustal melting causes highly discontinuous surface and mantle heat fluxes, the model outputs are the averaged values for the final 100 Myr of planetary evolution.} \label{fig:Mars_PCA}
\end{figure*}

Figures \ref{fig:Venus_PCA} and \ref{fig:Mars_PCA} summarize the results of 1320 simulations for Venus and Mars, respectively. Present-day values for selected output parameters are plotted against crustal thickness for both planets. Several correlations are readily apparent. For Venus, thicker crust is associated with higher Moho temperature, more mantle processing, higher mantle heat flux, and quicker crustal formation. More specifically, Moho temperature increases with crustal thickness in a linear fashion until $h_c \approx 75$ km, after which Moho temperatures remain near the critical value for basalt melting. Simulations with crustal melting have highly discontinuous surface and mantle heat fluxes, but such discontinuous nature is merely an artifact owing to our particular numerical implementation, so average values of $F_s$ and $F_m$ over the final 100 Myr of planetary evolution are used for all subsequent analyses. In contrast, Moho temperatures for Mars only approach the critical value for basalt melting in simulations with the thickest crust. Unlike for Venus, a decrease in present-day mantle heat flux accompanies an increase in crustal thickness for Mars.

Principal component analysis facilitates the interpretation of the correlations between output parameters. For Venus, two principal components account for most ($>$65$\%$) of the variance of the planetary parameters after 4.5 Gyr of thermal evolution. Calculations of the principle components returns coefficients with values between -1 and 1 that are associated with each model output parameter. Appendix A contains a table of principal component basis vectors for Venus. Comparing numerical values of select coefficients may reveal correlations with physical explanations. Arrows representing the eigenvectors associated with these principal components are plotted in Figs.~\ref{fig:Venus_PCA} and~\ref{fig:Mars_PCA}. These arrows indicate the axes along which the majority of the variance in the model output primarily lies. No preferred polarity exists for the principal component eigenvectors; plotting these arrows with a 180$^{\circ}$ rotation would be equally valid.

The first principal component represents the most dominant correlations among present-day planetary parameters, which are characterized mainly by the thicknesses of the crust and mantle lithosphere layers, as they are associated with large coefficients: $h_c$ (0.34), $h_l$ (-0.36), and $h_{ML}$ (-0.39). Because the sign of the coefficient for $h_c$ is opposite to the sign of the other two coefficients, the thicknesses of the crust and mantle lithosphere are anti-correlated. In other words, thick crust is associated with thin depleted mantle lithosphere and a thin thermal boundary layer and vice versa, since principal components have no preferred polarity. An initially hotter mantle produces thicker crust and thicker depleted lithosphere. Because mantle viscosity is lower for hotter mantle, however, the depleted lithosphere is more likely to be destabilized, resulting in a thinner lithosphere (and thus thermal boundary layer) for thicker crust. Other coefficients in the first principal component indicate the effects of crustal thickness on other model parameters, including the first-order correlations observed during inspection of Fig.~\ref{fig:Venus_PCA}. For instance, thick crust is associated with high Moho temperature and high surface and mantle heat fluxes. Thick crust also indicates a high degree of mantle processing and a corresponding low present-day mantle water content. Finally, the large negative coefficients for both $t_{c,10\%}$ and $\log_{10}(\Delta t_{c,tot})$ indicate that thick crust tends to form early and quickly.

The second principal component elucidates the effects of planet temperatures on other model parameters, since large coefficients are associated with $T_u$ (0.45) and $T_{cm}$ (0.44). Unsurprisingly, high mantle potential and core/mantle boundary temperatures are associated with high Moho temperature, since $T_c$ has a coefficient of 0.25. Moreover, high interior temperatures correspond to thick crust and a high degree of mantle processing, which would cause the present-day mantle water concentration to be very low. Note, however, that high (present-day) interior temperatures do not correspond to thick crust in case of the first principal component (Fig. 4b). In this space of crustal thickness and upper mantle temperature, the first and second principal components are nearly orthogonal, thus explaining the overall spread of simulation results. With principal component analysis, we can visualize how the most dominant trend (represented by the first principal component) is affected by secondary factors and how these secondary factors manifest in different parameter spaces. An important point is that the overall variability of planetary evolution can be compactly represented by a small number of principal components; that is, the effective dimension of the model space is actually small.

The principal components for Mars are very similar to those for Venus, with some notable exceptions. The first principal component again represents the effects of strongly correlated Moho temperature and crustal thickness, and thus explains the largest portion of the variance in the model output. As for Venus, a thin, cold crust is associated with thick depleted mantle lithosphere, a thick thermal boundary layer, a low surface heat flux, and a low degree of mantle processing. Unlike Venus, however, the surface and mantle heat fluxes in the first principal component are anti-correlated (see also Fig. \ref{fig:Mars_PCA}).

Despite the complexity of our thermal evolution model, some present-day parameters are found to be predicted with reasonable accuracy for Venus and Mars using a linear function.  A general formula for this function is
\begin{eqnarray}
B_i = A_{i,0} + A_{i,1}T_{u,n}(0) + A_{i,2}T_{cm,n}(0) + \nonumber \\
 A_{i,3} (\log_{10}(\eta_0))_n  +A_{i,4} (\log_{10}(\Delta\eta_w))_n + A_{i,5}Q_{0,n}, \label{eq:best-fit}
\end{eqnarray}
where $B_i$ is the value of the desired output parameter after 4.5 Gyr, constants $A_{i,0}$ through $A_{i,5}$ are estimated using the least-squares method for each $B_i$, and the subscript \textit{n} indicates that the input parameters are normalized and mean subtracted.

\begin{figure*}[ht]
\centering
\includegraphics[width=39pc]{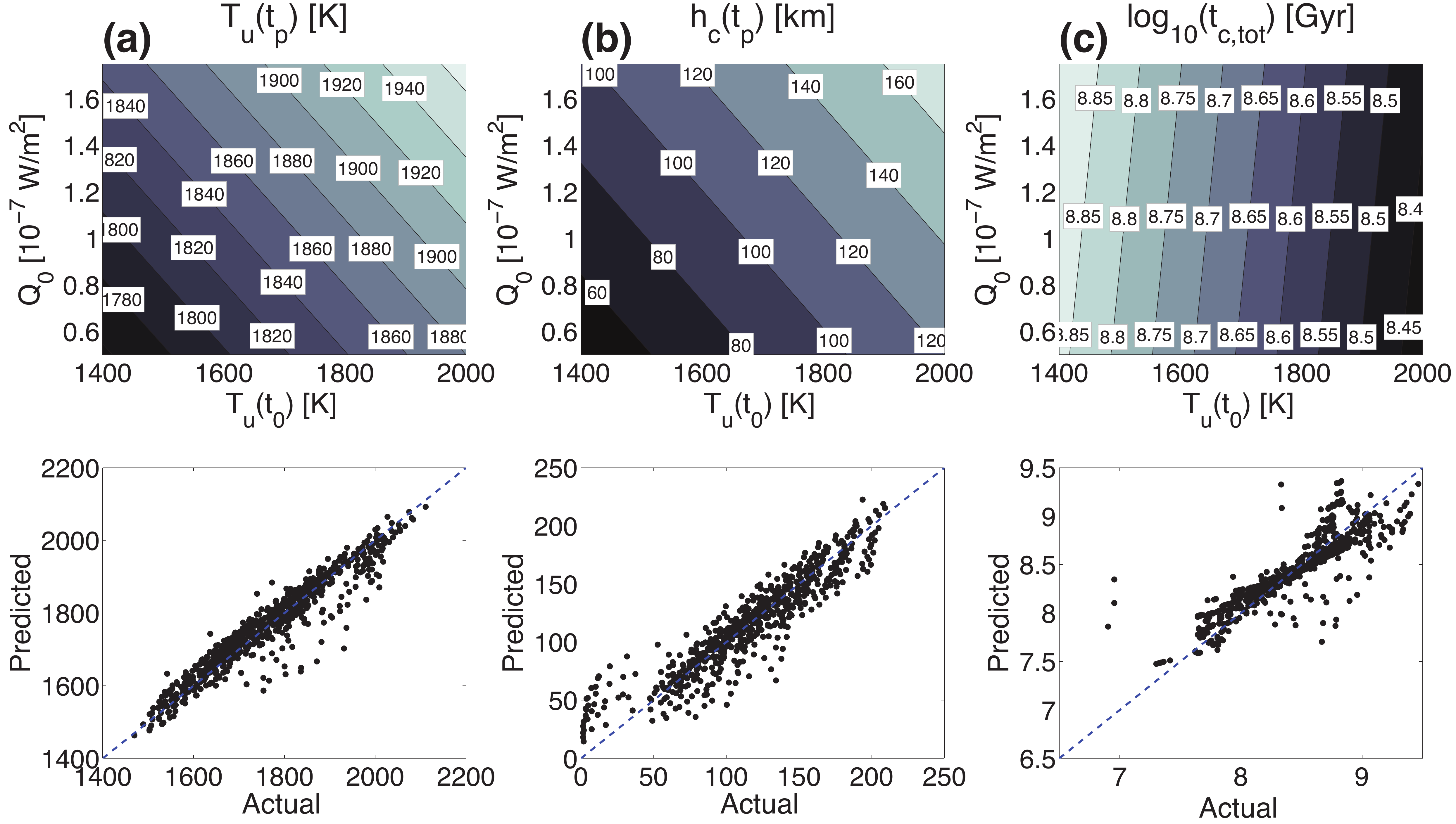}
\caption{Top panels show predicted values for Venus after 4.5 Gyr given initial volumetric radiogenic heating and initial mantle potential temperature, the initial conditions to which the output is most sensitive. Bottom panels show the correspondence between predicted and actual simulation results. The dashed line represents perfect predictive power. Default initial conditions are $T_{cm}(0) = 4000$ $\text{K}$, $\eta_0 = 10^{19}$ $\text{Pa}\cdot\text{s}$, $\Delta\eta_w=100$, and $(d\rho/d\phi) = 120$ kg m$^{-3}$. Panels show predicted values of (a) mantle potential temperature, (b) crustal thickness, and (c) total time for crust to grow from 10\% to 95\% of its present thickness.} \label{fig:Venus_bestfit}
\end{figure*}

Table~\ref{table:Venus_bestfit} lists constants for Venusian $B_i$ that have relatively high correlation coefficients between predicted and actual simulation results. Figure~\ref{fig:Venus_bestfit} shows contour plots with predicted values of mantle potential temperature, crustal thickness, and duration of crustal formation for given initial internal heating and mantle potential temperature. While present-day mantle potential temperature and crustal thickness depend strongly on both initial mantle potential temperature and the magnitude of internal heating, the total duration of crustal formation is primarily a function of initial mantle potential temperature (see Table~\ref{table:Venus_bestfit} for more complete information on parameter sensitivity). Figure~\ref{fig:Venus_bestfit} also demonstrates a reasonable correspondence between the predicted and actual values of these model outputs for all of the simulations. This way of summarizing simulation results allows us not only to see the sensitivity of model outputs to initial parameters but also to quickly reproduce major modeling results without redoing simulation.

\begin{table*}[ht]
\begin{center}
\begin{tabular}{c*{6}{c}*{2}{c}}
\hline
B$_i$ & A$_{i,0}$ & A$_{i,1}$ & A$_{i,2}$ & A$_{i,3}$ & A$_{i,4}$ & A$_{i,5}$ & Units & Corr. \\ \hline
$T_u$ &  1772 & 43.6 & 11.3 & 89.9 & 69.4 & 30.4 & K & 0.95 \\
$T_{cm}$ &  3128 & 88.5 & 32.5 & 137.0 & 104.8 & 48.8 & K & 0.94 \\
$h_c$ & 115 & 26.4 & 22.2 & -2.98 & -7.84 & 18.0 & km & 0.92 \\
$h_l$ &  53.6 & -2.17 & -5.51 & 17.5 & 17.8 & -6.07 & km & 0.91 \\
$h_{ML}$ & 84.1 & -4.82 & -4.87 & 19.2 & 14.9 & -6.87 & km & 0.94 \\
$F_s$ & 50.7 & 3.96 & 2.38 & -1.87 & -0.76 & 13.5 & mW m$^{-3}$ & 0.98 \\
$F_m$ & 29.6 & 3.40 & 1.51 & -1.91 & -1.55 & 2.92 & mW m$^{-3}$ & 0.81 \\
log$_{10}(u$) & 0.89 & 0.11 & 0.01 & -0.15 & -0.07 & 0.04 & - & 0.82 \\
$V_{proc}/V_{sm}$ & 1.04 & 0.00 & 0.06 & -0.01 & 0.07 & 0.16 & - & 0.74 \\
$\log_{10}(\Delta{t_{c,tot}})$ & 8.42 & -0.15 & -0.16 & 0.22 & 0.20 & 0.01& -  & 0.87 \\ \hline
\end{tabular}
\end{center}
\caption{Coefficients for the best-fit linear function (Eq.~\ref{eq:best-fit}) relating parameter values after 4.5 Gyr for parameters with correlation coefficients $>$ 0.70 to a given set of initial conditions for Venus. Correlation coefficients quantifying the correspondence between the actual and predicted output parameters were calculated using normalized and mean subtracted input and output parameters. The average values of the input parameters are $T_u(0)$ = 1700 K, $T_{cm}(0)$ = 4000 K, $\log_{10}(\eta_0)$ = 19, $\log_{10}(\Delta\eta_w)$ = 1, and $Q_0$ = 1.05 $\times$ 10$^{-7}$ W m$^{-3}$. For the best-fit function, the input parameters are mean subtracted and normalized by 212 K, 408 K, 0.82, 0.82, and 4.30 $\times$ 10$^{-8}$ W m$^{-3}$, respectively.} \label{table:Venus_bestfit} 
\end{table*}

\subsection{Evolution of Super-Venus Planets}

We investigate the evolution of super-Venus planets to explore the effects of planetary mass on stagnant-lid convection. For simplicity, surface temperatures for all super-Venus planets are assumed to be 300 K, though in reality this temperature may vary with time and is highly dependent on atmospheric composition and on the distance to and the luminosity of the central star. 

\subsubsection{Sample Thermal Histories}

\begin{figure*}[ht]
\centering
\includegraphics[width=39pc]{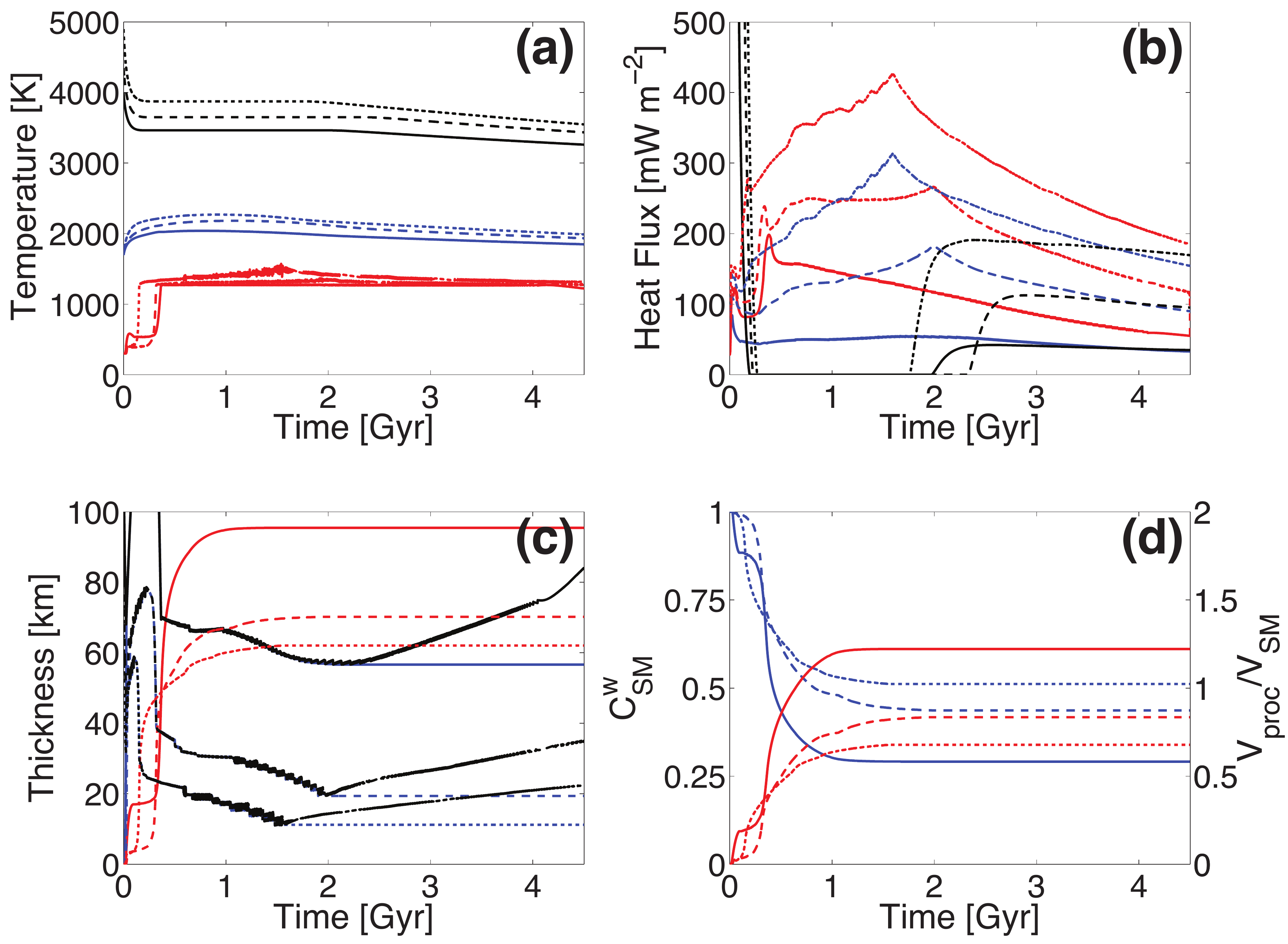}
\caption{Sample histories for 1$M_\oplus$ (solid lines), 5$M_\oplus$ (dashed lines), and 10$M_\oplus$ (dotted lines) super-Venus planets. Red, blue, and black curves, respectively, signify (a) crust, mantle potential, and core/mantle boundary temperatures, (b) surface, mantle, and core heat flows, (c) crust, depleted mantle lithosphere, and mantle lithosphere thicknesses, and (d) normalized mantle water content and fraction of processed source mantle. Default initial conditions are $Q_0$ = 1.0 $\times10^{-7}$ W m$^{-3}$ (scaled with $\rho_m$), $T_u(0) = 1700$ K, $\eta_0 = 10^{19}$ $\text{Pa}\cdot\text{s}$, $\Delta\eta_w=100$, and $(d\rho/d\phi) = 120$ kg m$^{-3}$. The 1, 5, and 10$M_\oplus$ planets have $T_{cm}(0) =$ 4000, 4350, and 4900 K, respectively. Because crustal melting causes highly discontinuous surface and mantle heat fluxes, a moving average with a 75 Myr span was used for plotting purposes.} \label{fig:superHist}
\end{figure*}

Super-Venus planets with $M_P$ = 1, 5, and 10$M_\oplus$ were evolved to study the effects of increasing planetary mass on a variety of parameters, particularly crustal production. For all three planets, $Q_0$ was scaled to the Venusian value of 1.0$\times10^{-7}$ W m$^{-3}$, $T_u(0)$ = 1700 K, and $T_{cm}(0)$ = 4000, 4350, and 4900 K, respectively. Dehydration stiffening and compositional buoyancy were both incorporated as usual. Figure~\ref{fig:superHist} shows the results of these simulations. As with Venus and Mars, the transient ``hot start" in the core is lost in the first $\sim$100 Myr. After this initial cooling, mantle dynamics controls core cooling. Because the mantle heats up for the first $\sim$1 Gyr and then cools only very slowly, core cooling is precluded for the first $\sim$2 Gyr. As suggested by simple scaling laws \citep{DJS2003CRG}, mantle cooling paths for massive super-Venus planets are roughly parallel.

Figure~\ref{fig:superHist} also shows how the thicknesses of the crust, mantle lithosphere, and depleted mantle lithosphere vary with time. With increasing planetary mass, crustal thickness decreases. The simple scaling analyses below indicate that more massive planets have greater melt production. The observed increase in mantle potential temperature with planetary mass only accentuates this effect. The increased melt volume, however, is not sufficient to create a thicker crust on a larger planet. The 1$M_\oplus$ planet in the stagnant-lid regime ceases crustal production soon after 1 Gyr as mantle potential temperature drops below a critical value. The increased interior temperatures for the more massive planets allow longer durations of crustal production. For the first $\sim$2 Gyr of thermal evolution, the thickness of the depleted mantle lithosphere is close to that of the mantle thermal lithosphere, reflecting the continuous delamination of excess depleted mantle lithosphere. Decreased crustal production with increasing planetary mass corresponds to a smaller degree of mantle processing and a higher content of residual mantle water.
 
\subsubsection{Sensitivity Analyses}

\begin{figure*}[ht]
\centering
\includegraphics[width=39pc]{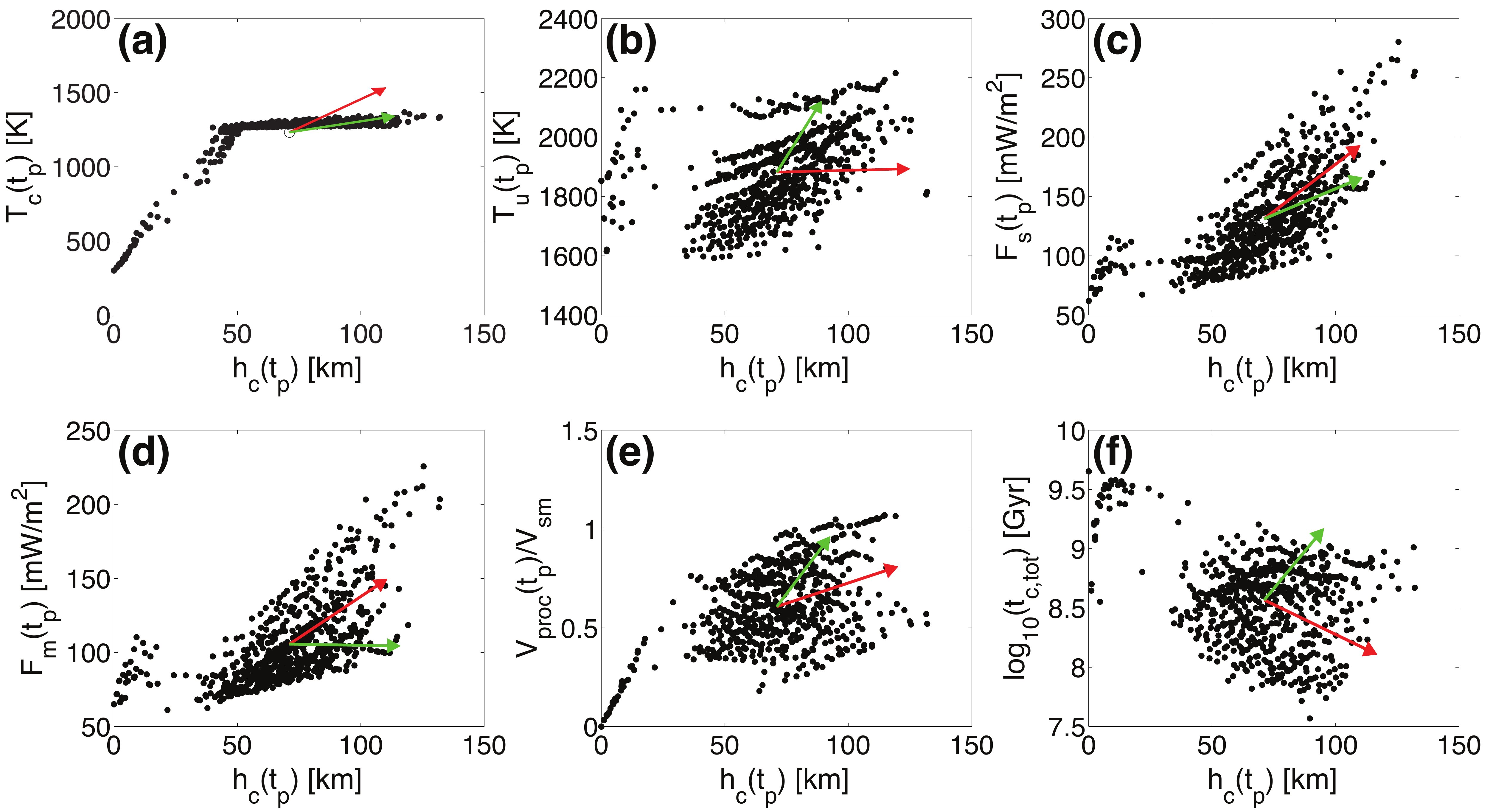}
\caption{Summary of parameter values at the present for 672 simulations of the thermal evolution of a $5M_\oplus$ super-Venus. Arrows are projections of the principal component basis vectors that emanate from a point representing the averaged simulation results, indicating axes that account for the vast majority of the data set's variance. The red arrow represents a larger percentage of cumulative variance (41\%) than the green arrow (28\%). Panels show (a) Moho temperature, (b) mantle potential temperature, (c) surface heat flux, (d) mantle heat flux, (e) fraction of mantle processed by melting, and (f) total time for crust to grow from 10\% to 95\% of its present thickness as functions of crustal thickness. Because crustal melting causes highly discontinuous surface and mantle heat fluxes, the model output is the averaged values for the final 100 Myr of planetary evolution.} \label{fig:5Me_PCA}
\end{figure*}

\begin{figure*}[ht]
\centering
\includegraphics[width=39pc]{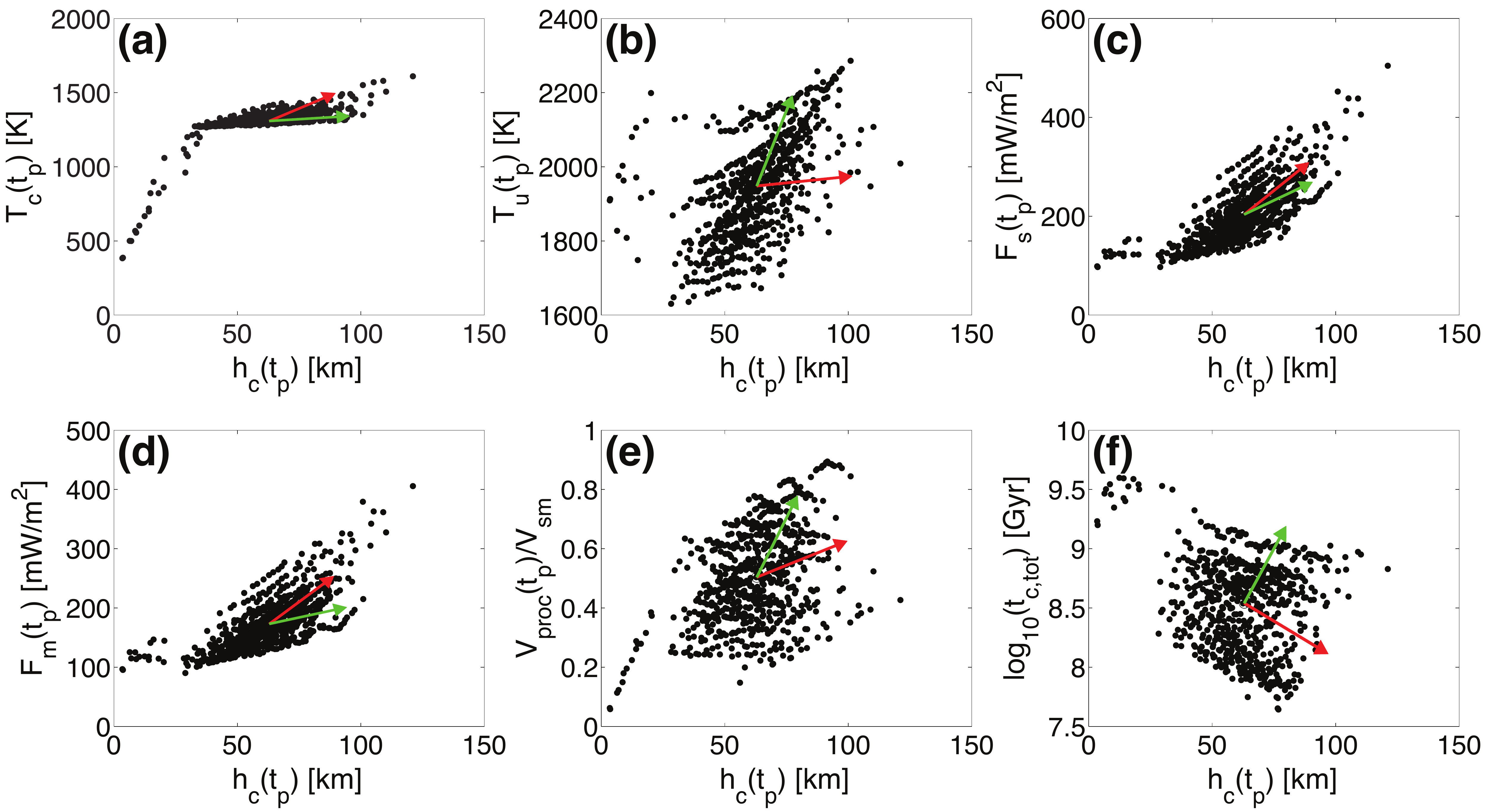}
\caption{Summary of parameter values at the present for 675 simulations of the thermal evolution of a $10M_\oplus$ super-Venus. Arrows are projections of the principal component basis vectors that emanate from a point representing the averaged simulation results, indicating axes that account for the vast majority of the data set's variance. The red arrow represents a larger percentage of cumulative variance (42\%) than the green arrow (30\%). Panels show (a) Moho temperature, (b) mantle potential temperature, (c) surface heat flux, (d) mantle heat flux, (e) fraction of mantle processed by melting, and (f) total time for crust to grow from 10\% to 95\% of its present thickness as functions of crustal thickness. Because crustal melting causes highly discontinuous surface and mantle heat fluxes, the model output is the averaged values for the final 100 Myr of planetary evolution.} \label{fig:10Me_PCA}
\end{figure*}

The output of 1347 simulations for 5 and 10$M_\oplus$ super-Venus planets are shown in Figs.~\ref{fig:5Me_PCA} and \ref{fig:10Me_PCA}. Three simulations for the 5$M_\oplus$ super-Venus planet were excluded because they did not meet the requirements that $h_c < 500$ km and that inner core growth did not occur. As for Mars and Venus, present-day parameters of interest are plotted against present-day crustal thickness. The principal component eigenvectors, explained below, are projected onto each plot, emanating from the average simulation output. The table in the appendix contains the principal component basis vectors for the 10$M_\oplus$ planet.

These scatter plots reveal similarities between the evolution of both massive planets. For instance, Moho temperature increases with crustal thickness in a linear fashion before reaching the critical value for basalt melting. With increasing planetary mass, the critical crustal thickness at which this transition occurs decreases. For relatively thick crust, Moho temperatures remain near the critical value for basalt melting. For both super-Venus planets, an increase in crustal thickness is associated with an increase in present-day mantle potential temperature, mantle heat flux, and degree of mantle processing. The total duration of crustal formation decreases with increasing present-day crustal thickness. Again, correlations between model parameters may be studied in more detail with principal component analysis.

For both planets, as for Venus and Mars, the first principal component is characterized by a strong correlation between Moho temperature and crustal thickness, explaining the general trends observed in Figs.~\ref{fig:5Me_PCA} and \ref{fig:10Me_PCA}. A decrease in both quantities is associated with an increase in the thicknesses of the depleted mantle lithosphere and the thermal boundary layer, a decrease in surface and mantle heat fluxes, an increase in the duration of crustal formation, and a decrease in the degree of mantle processing. The second principal component illuminates the effect of correlated interior temperatures. As expected, increasing mantle potential and core/mantle boundary temperatures causes an increases in crustal thickness, the total duration of crustal formation, and the degree of mantle processing. 

\begin{table*}[ht]
\begin{center}
\begin{tabular}{c*{6}{r}*{2}{c}}
\hline
B$_i$ & A$_{i,0}$ & A$_{i,1}$ & A$_{i,2}$ & A$_{i,3}$ & A$_{i,4}$ & A$_{i,5}$ & Units & Corr. \\ \hline
$T_u$ &  1882 & 42.1 & 13.6 & 93.9 & 57.7 & 46.6 & K & 0.92 \\
$T_{cm}$ &  3331 & 94.1 & 41.7 & 132.2 & 79.0 & 74.7 & K & 0.92 \\
$h_c$ & 71.1 & 12.4 & 8.31 & 1.47 & -0.64 & 14.6 & km & 0.89 \\
$h_l$ &  17.5 & -1.31 & -1.71 & 5.35 & 4.87 & -3.97 & km & 0.74 \\
$h_{ML}$ & 32.0 & -2.89 & -2.42 & 7.57 & 5.00 & -5.08 & km & 0.84 \\
$F_s$ & 131.1 & 12.9 & 7.56 & -8.81 & -4.12 & 32.1 & mW m$^{-3}$ & 0.95 \\
$F_m$ & 105.6 & 11.8 & 6.57 & -9.08 & -6.87 & 17.7 & mW m$^{-3}$ & 0.89 \\
log$_{10}(u$) & 1.65 & 0.10 & 0.03 & -0.18 & -0.07 & 0.07 & - & 0.84 \\
$C^w_{sm}$ & 0.57 & 0.00 & -0.01 & -0.01 & -0.06 & -0.07 & - & 0.78 \\
$V_{proc}/V_{sm}$ & 0.61 & 0.00 & 0.02 & 0.02 & 0.11 & 0.13 & - & 0.83 \\
$t_{c,10\%}$ & 0.20 & -0.15 & -0.10 & 0.10 & 0.08 & -0.03 & Gyr & 0.76 \\
$\log_{10}(\Delta{t_{c,tot}})$ & 8.57 & -0.09 & -0.12 & 0.17 & 0.19 & 0.03 & -  & 0.79 \\ \hline
\end{tabular}
\end{center}
\caption{Coefficients for the best-fit linear function (Eq.~\ref{eq:best-fit}) relating parameter values after 4.5 Gyr for parameters with correlation coefficients $>$ 0.70 to a given set of initial conditions for a 5$M_\oplus$ super-Venus. Correlation coefficients quantifying the correspondence between the actual and predicted output parameters were calculated using normalized and mean subtracted input and output parameters. The average values of the input parameters are $T_u(0)$ = 1701 K, $T_{cm}(0)$ = 4351 K, $\log_{10}(\eta_0)$ = 19, $\log_{10}(\Delta\eta_w)$ = 1, and $Q_0$ = 1.73 $\times$ 10$^{-7}$ W m$^{-3}$. For the best-fit function, the input parameters are mean subtracted and normalized by 212 K, 408 K, 0.82, 0.82, and 7.07 $\times$ 10$^{-8}$ W m$^{-3}$, respectively.} \label{table:5Me_bestfit} 
\end{table*}

Many present-day model parameters of interest can be represented as a linear function of initial conditions (Table~\ref{table:5Me_bestfit}). Compared to the case of Venus, a greater number of parameters are found to be approximated reasonably well by this approach. The effects of melting at the base of the crust undoubtedly remain a large source of nonlinearity in the model output for all terrestrial planets more massive than Mars. A more elaborate numerical implementation to deal with exceedingly high crustal temperatures may reduce such nonlinearity, though we did not explore this possibility.

\subsection{Scaling of Crustal Thickness and Mantle Processing}

We conduct simple scaling analyses to better understand the cause of decreasing crustal thickness and a decreasing degree of mantle processing with increasing planetary mass.

\subsubsection{Crustal Thickness}

A number of parameters govern the scaling of crustal thickness with planetary mass. Increased melt production, for instance, is the first requirement for thicker crust. From Eq. \ref{eq:melt}, volumetric melt production for a planet may scale as
\begin{equation}
\frac{f_m}{f_{m,\oplus}} = \left(\frac{d_m}{d_{m,\oplus}}\right)\left(\frac{u}{u_{\oplus}}\right) \left(\frac{\phi}{\phi_\oplus}\right) \left(\frac{h_{m,\oplus}}{h_m}\right) \left(\frac{A_m}{A_{m,\oplus}}\right) \approx \left(\frac{M}{M_\oplus}\right)^\delta,
\end{equation}
where the subscript $\oplus$ denotes values for an Earth-mass planet and $A_m$ stands for the mantle surface area.

We can approximate $\delta$ using the representative interior models of \citet{V06}, for which $R\propto M^{0.262}$, $\rho_m\propto M^{0.196}$, and $g\propto M^{0.503}$. First, consider the thickness of a melting region, $d_m$ = $z_i$ - $z_f$. Since $z_f$ = $P_f/(\rho_Lg)$ is approximately constant for any planet,
\begin{equation} 
\frac{d_m}{d_{m,\oplus}} \approx \frac{g_\oplus}{g} = \left(\frac{M}{M_\oplus}\right)^{-0.503},
\end{equation}
where a roughly constant mantle to core thickness ratio is assumed, although planetary mantles grow slightly more than cores with increasing planetary mass.
Next, 
\begin{equation} 
\frac{u}{u_{\oplus}} = \frac{h_{m,\oplus}}{h_m}\left(\frac{Ra}{Ra_\oplus}\right)^{\frac{1}{2}}.
\end{equation}
The Rayleigh number for a massive planet scales as
\begin{equation}
\frac{Ra}{Ra_{\oplus}} = \left(\frac{\Delta T_u}{\Delta T_{u,\oplus}}\right)\left(\frac{\eta(T_u)}{\eta(T_{u,\oplus})}\right)\left(\frac{g}{g_\oplus}\right)\left(\frac{\rho}{\rho_\oplus}\right)\left(\frac{h_m}{h_{m,\oplus}}\right)^{3}.
\end{equation}
Assuming that the first and second terms on the right hand side are roughly equal to unity, we have
\begin{equation}
\frac{Ra}{Ra_{\oplus}} \approx \left(\frac{M}{M_\oplus}\right)^{1.485}
\end{equation}
and thus
\begin{equation}
\frac{u}{u_{\oplus}} \approx \left(\frac{M}{M_\oplus}\right)^{0.481}.
\end{equation}
Because $h_c$ is usually much smaller than $R_P$,
\begin{equation} 
\frac{A_m}{A_{m,\oplus}} = \left(\frac{R_P-h_c}{R_{P,\oplus}-h_{c,\oplus}}\right)^2 \approx \left(\frac{R_P}{R_{P,\oplus}}\right)^2 =  \left(\frac{M}{M_\oplus}\right)^{0.524}.
\end{equation}
Finally, the rest of the scaling relations may simply be assumed as
\begin{equation} 
\frac{\phi}{\phi_\oplus} \approx 1 
\end{equation}
and
\begin{equation}
\frac{h_{m,\oplus}}{h_m} \approx \left(\frac{M}{M_\oplus}\right)^{-0.262}. 
\end{equation}

Hence, $\delta \approx 0.240$ and $(f_m/f_{m,\oplus})\approx (M/M_\oplus)^{0.240}$. Because $h_c$ $\approx$ $f_m \times \Delta t/(4\pi R_P^2)$, where $\Delta t$ is the duration of crust growth, an increase in melt productivity with mass does not guarantee an increase in crustal thickness with mass. As planetary mass, and thus radius, increases, a larger volumetric melt production is required to produce a certain crustal thickness. Specifically, crustal thickness would only increase with mass for $\delta > 0.524$, assuming that $\Delta t$ is roughly constant. Therefore, although melt productivity increases with planetary mass, this simple scaling analysis indicates that crustal thickness should decrease with scaling $(h_c/h_{c,\oplus})\approx (M/M_\oplus)^{(0.240-0.524)}$ $=(M/M_\oplus)^{-0.284}$~. 

\begin{figure}
\centering
\includegraphics[width=20pc]{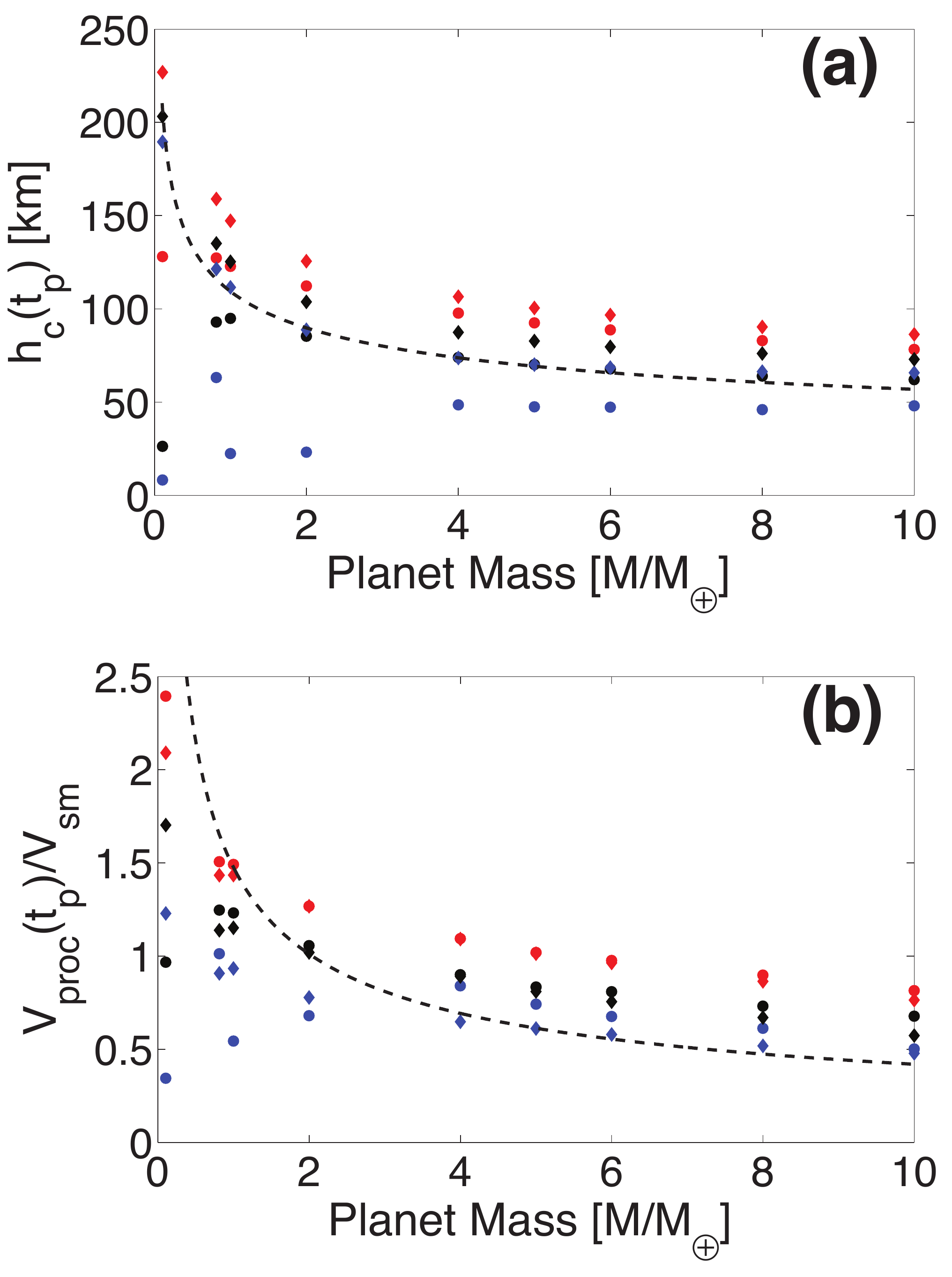}
\caption{Summary of 54 simulations of the evolution of Mars, Venus, and seven super-Venus planets, showing the correspondence between simulation results and simple scaling laws for the effects of planetary mass on (a) crustal thickness and (b) mantle processing. Circles and triangles represent $T_u(0)$ = 1700 and 2000 K, respectively. Blue, black, and red symbols represent Venus-equivalent $Q_0$ = $5.0\times10^{-8}$, $1.0\times10^{-7}$, and $1.75\times10^{-7}$ W m$^{-3}$, respectively. Dashed black lines show the scaling relations (a) $h_c \propto (M/M_\oplus)^{-0.284}$ and (b) $(V_{proc}/V) \propto (M/M_\oplus)^{-0.546}$, with each curve fixed to intersect the average output from the simulations for the 2$M_\oplus$ super-Venus planet.} \label{fig:scaling}
\end{figure}

Panel (a) of Fig.~\ref{fig:scaling} is a plot of model output present-day crustal thickness as a function of planetary mass for simulations of Mars, Venus, and seven super-Venus planets. While initial conditions strongly affect simulation results, the model outputs generally follow this simple scaling. Smaller planets can have thicker crust though they tend to be characterized by lower mantle temperatures.

\subsubsection{Mantle Processing}

The scaling of mantle processing with planetary mass follows easily from the above analysis. A simplified equation for the volume of processed mantle is
\begin{equation}
V_{proc} \approx \frac{f_m}{\phi}\Delta t,
\end{equation}
where $\Delta t$ is a duration for crustal growth.

Thus, the amount of processed mantle scales with planetary mass as
\begin{equation}
\frac{V_{proc}}{V_{proc,\oplus}} \approx \left(\frac{f_m}{f_{m,\oplus}}\right) \left(\frac{\phi_\oplus}{\phi}\right) \approx \left(\frac{M}{M_\oplus}\right)^\xi,
\end{equation}
so $\xi \approx \delta \approx 0.240$.

The volume of a super-Venus planet scales as
\begin{equation}
\frac{V}{V_\oplus} = \left(\frac{R}{R_\oplus}\right)^3 \approx \left(\frac{M}{M_\oplus}\right)^\zeta, 
\end{equation}
so $\zeta=0.786$. Therefore, $(V_{proc}/V)$ $\propto (M/M_\oplus)^{-0.546}$. Although the amount of processed mantle material increases with planetary mass, the fraction of processed mantle decreases with increasing planetary mass because the mantle volume increases more rapidly than the amount of processed material. Panel (b) in Fig.~\ref{fig:scaling} confirms that the fraction of processed mantle does indeed decrease with increasing planetary mass according to this scaling law, although initial conditions strongly affect the simulation results.

\subsection{Viscosity Contrasts During Stagnant-Lid Convection}

\begin{figure}
\centering
\includegraphics[width=20pc]{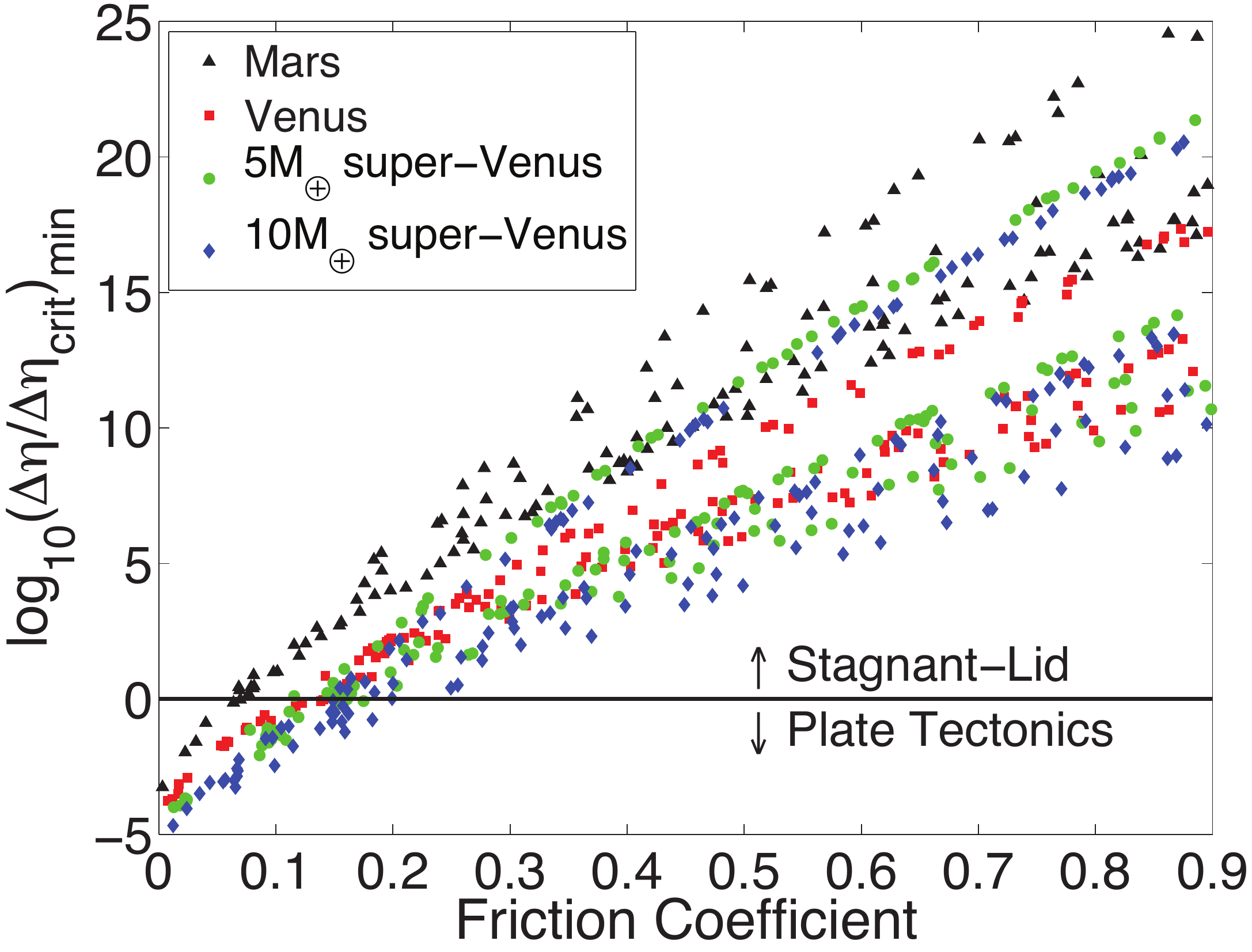}
\caption{Summary of 595 simulations of the evolution of Mars, Venus, and two super-Venus planets, showing the minimum ratio of actual viscosity contrast to critical viscosity contrast and thus the likelihood of plate tectonics being favored at some point during 4.5 Gyr of planetary evolution.  Each planet was evolved from six different sets of initial conditions (three values for both radiogenic heating and mantle potential temperature) for many different values of $\mu$, the effective friction coefficient. Points plotted above the indicated line represent simulations for which the actual viscosity contrast never dipped below the critical value for a transition to plate tectonics. Below the indicated line, which occurs only for $\mu < 0.3$, plate tectonics may have been favored at some point. For dry silicate rocks, $\mu\sim$ 0.7 to 0.8.} \label{fig:tectonics}
\end{figure}

The viscosity contrast across the lithosphere is tracked during each thermal evolution simulation, along with the critical viscosity contrast above which a planet is locked in the stagnant-lid regime. Figure~\ref{fig:tectonics} shows the output of 595 simulations for Mars, Venus, and two super-Venus planets for which $Q_0$, $T_u(0)$, and $\mu$ were varied over a wide range. In particular, all permutations of $Q_0$ = 0.5, 1.0, and 1.75 $\times10^{-7}$ W m$^{-3}$ (scaled as usual with $\rho_m$) and $T_u(0)$ = 1400, 1700, and 2000 K were considered for a range of $\mu$ between 0.0 and 0.9. 

From this plot, several conclusions may be drawn. First, for values of the frictional coefficient associated with dry silicate rocks, $\mu \sim$ 0.7 to 0.8, plate tectonics is never favored. Second, increasing planetary mass does not substantially affect the likelihood of plate tectonics. Third, the effects of choosing different initial conditions are amplified for greater planetary mass. Finally, although choosing extreme initial conditions can change the viscosity contrast by orders of magnitude, the effect of the friction coefficient is far more important.

\subsection{Formation of an Eclogite Layer}

At depth, crustal rock may undergo a phase transition to eclogite. To extend the simple analysis from earlier, we write the thickness of the crust in the eclogite stability field as $h_e$ = $h_c$ - $d_e$, where $d_e$ is the depth of the phase boundary. Likewise, we consider $\delta_e$ and $\delta_a$, the fractions of the crust in and above, respectively, the eclogite stability field. Because $\delta_e + \delta_a =  \delta_{e,\oplus} + \delta_{a,\oplus} = 1$, we may write
\begin{equation}
\delta_e=\delta_{e,\oplus}+\delta_{a,\oplus}\left(1-\frac{\delta_a}{\delta_{a,\oplus}}\right).
\end{equation}
The fraction of the crust above the eclogite stability field may scale as
\begin{equation}
\frac{\delta_a}{\delta_{a,\oplus}}=\frac{d_e/h_c}{d_{e,\oplus}/h_{c,\oplus}}=\left(\frac{h_{c,\oplus}}{h_c}\right)\left(\frac{d_e}{d_{e,\oplus}}\right)\approx\left(\frac{M}{M_\oplus}\right)^\epsilon.
\end{equation}
If we assume that pressure increases hydrostatically with depth and that the critical pressure below which the phase transition occurs is a constant, then $d_e \sim 1/g$. So, $\epsilon$ = 0.284 - 0.503 = -0.219. Therefore, the fraction of crust in the eclogite stability field should increase with planetary mass as
\begin{equation}
\delta_e=\delta_{e,\oplus}+\delta_{a,\oplus}\left[1-\left(\frac{M}{M_{\oplus}}\right)^{-0.219}\right]. \label{eq:eclogite}
\end{equation}

In thermal evolution models, the heat conduction equation is numerically solved to calculate crustal temperatures. An approximate temperature profile can also be calculated using a steady-state approximation as \citep{TS2002}
\begin{equation}
\label{eq:temp_profile}
T(z) = T_s + \frac{F_s}{k}z - \frac{Q_c(t_p)}{2k}z^2,
\end{equation}
where $Q_c$, the volumetric crustal heat production, is calculated as
\begin{equation}
Q_c(t_p) = Q_0e^{-\lambda t_p}\left(\frac{V_{proc}(t_p)}{V_{c}(t_p)}\right), 
\end{equation}
where $t_p$ is 4.5 Gyr and $V_c$ is the volume of the crust. The boundary condition $T(h_c) = T_c$ is used to calculate surface heat flux for specified Moho and surface temperatures and magnitude of internal heat production. Finally, Eq.~\ref{eq:temp_profile} is used to calculate the temperature profile throughout the entire thickness of the crust. Representative temperature profiles for a planet can be used to approximate the fraction of crust that lies within the eclogite stability field.

\begin{figure*}[ht]
\centering
\includegraphics[width=39pc]{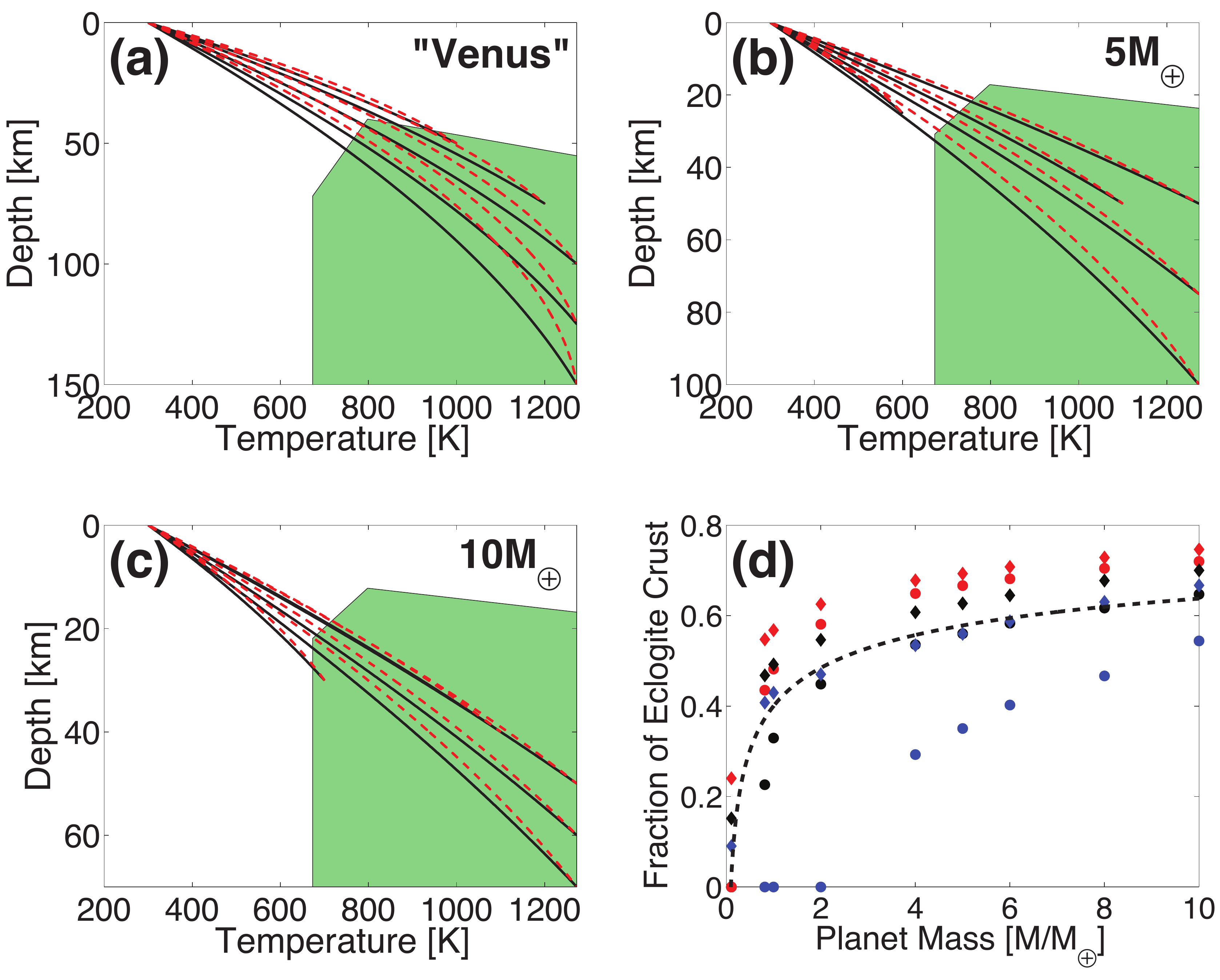}
\caption{Crustal temperature profiles for (a) Venus and (b) 5$M_\oplus$ and (c) 10$M_\oplus$ super-Venus planets, calculated assuming representative crustal thicknesses, degrees of mantle processing, and Moho temperatures. The green shaded area is the approximate stability field for eclogite, drawn using the phase diagram from \citet{POMBook} and assuming a hydrostatic pressure increase with depth. Black solid and red dashed lines represent Venus-equivalent $Q_0$ = $1.0\times10^{-7}$ and $1.75\times10^{-7}$ W m$^{-3}$, respectively. Panel (d) shows the fraction of crust in the eclogite phase for Mars, Venus, and seven super-Venus planets. Circles and triangles represent $T_u(0)$ = 1700 and 2000 K, respectively. Blue, black, and red symbols represent Venus-equivalent $Q_0$ = $5.0\times10^{-8}$, $1.0\times10^{-7}$, and $1.75\times10^{-7}$ W m$^{-3}$, respectively.} \label{fig:eclogite}
\end{figure*}

Figure~\ref{fig:eclogite} shows representative temperature profiles for Venus and 5 and 10$M_\oplus$ super-Venus planets, calculated using representative crustal thicknesses, degrees of mantle processing, and Moho temperatures from the previous sensitivity analyses. A range of internal radiogenic heating was also considered. The stability field of eclogite is taken from \citet{POMBook} and is drawn assuming a hydrostatic pressure increase with depth. Panel (d) in Fig.~\ref{fig:eclogite} summarizes the effects of initial conditions on the fraction of crust in the eclogite stability region after 4.5 Gyr and shows the scaling from Eq.~\ref{eq:eclogite}. For Mars, Venus, and seven super-Venus planets, 54 thermal evolution simulations were run to study all permutations of the initial conditions $T_u(0)$ = 1700 and 2000 K and Venus-equivalent $Q_0$ = 0.5, 1.0, and 1.75 $\times10^{-7}$ W m$^{-3}$. As predicted, the fraction of eclogite crust increases with planetary mass.

\section{Discussion}

\subsection{Pressure Effects on Mantle Rheology}

The rheological behavior of the mantles of large rocky planets is difficult to predict. While the core/mantle boundary pressure for Earth is $\sim$135 GPa \citep[e.g.,][]{PREM}, pressures within the silicate mantles of large rocky planets likely exceed 1~TPa \citep[e.g.,][]{V06}. Above the transition to post-perovskite at $\sim$120~GPa, Earth's mantle is primarily made of MgSiO$_3$-perovskite and MgO \citep[e.g.,][]{Murakami2004}. In contrast, much of the mantles of super-Venus planets will be dominated by post-perovskite and perhaps, above 1~TPa, a mixture of MgO and SiO$_2$ \citep{Umemoto2006}. Unfortunately, we lack experimental measurements of the properties of planetary materials under these extreme conditions. Until such data are available, conjectures about the rheology of the silicate mantles of large rocky planets will remain controversial. Thermal evolution simulations are very sensitive to assumed rheological behaviors, so investigating the implications of various possible assumptions is essential.

\begin{figure}
\centering
\includegraphics[width=20pc]{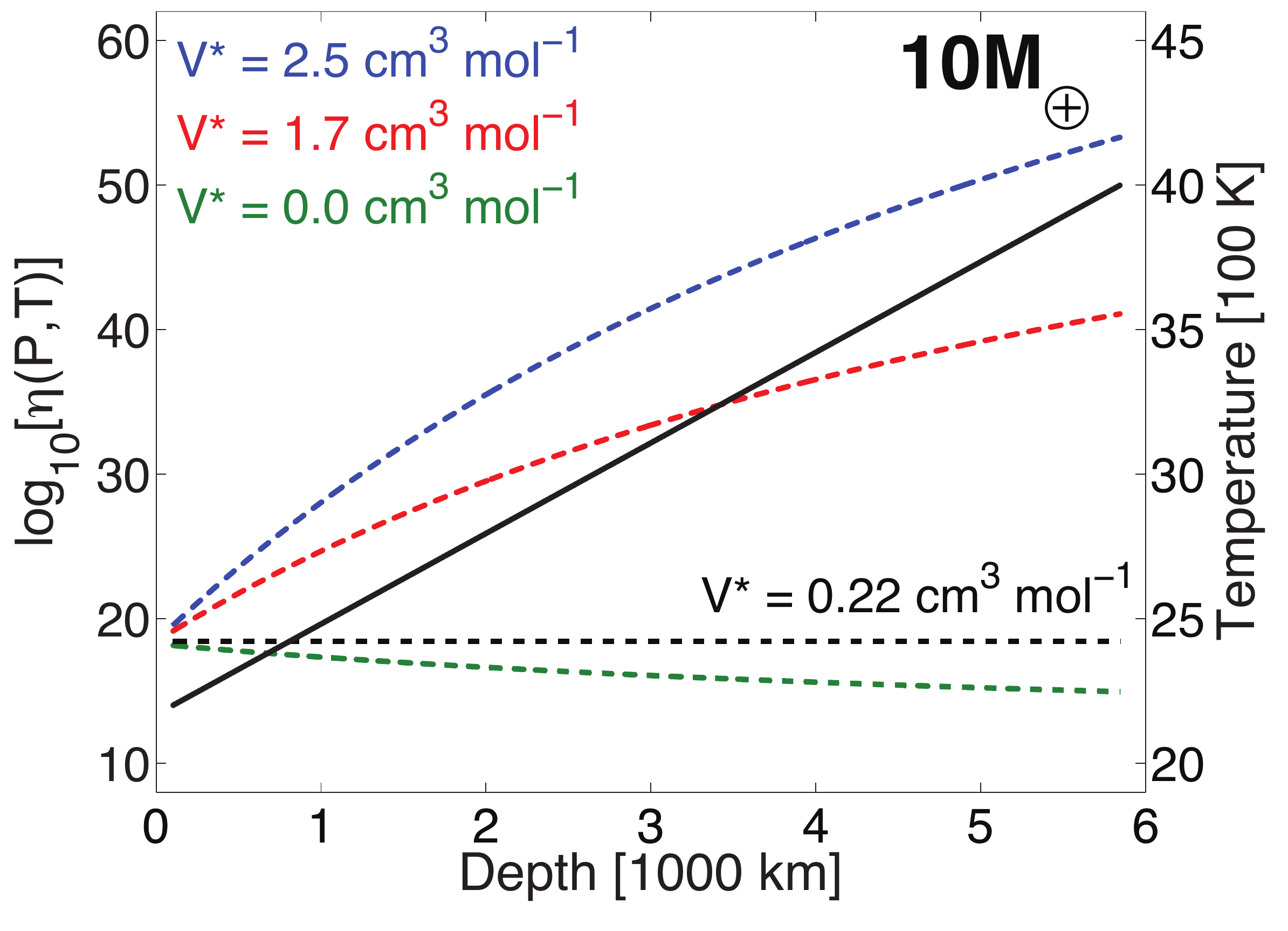}
\caption{Internal temperature and viscosity profiles for a 10$M_\oplus$ super-Venus planet, calculated as in \citet{Vlada2012}. The black curve shows internal temperature as a function of depth in the convecting mantle. Green, red, and blue dashed lines represent viscosity profiles calculated using Eq.~\ref{eq:viscS} for $V^*$ = 0, 0.22, 1.7, and 2.5 cm$^3$ mol$^{-1}$, respectively.} \label{fig:visc}
\end{figure}

The viscosities of most planetary materials increase with pressure when examined at relatively low pressures \citep[e.g.,][]{Karato2008}. Simple extrapolation of this trend predicts extreme increases in viscosity within massive terrestrial planets. Extensions of known perovskite rheology, for instance, imply an increase of $>$15 orders of magnitude as pressure increases to 1~TPa in an adiabatic mantle \citep{Vlada2011}. Specifically, a viscosity profile may be calculated as \citep{Vlada2012}
\begin{equation}
\eta(P,T) = \eta_0\exp\left[\frac{E}{R}\left(\frac{1}{T}-\frac{1}{T^*}\right)+\frac{1}{R}\left(\frac{P V^*}{T}\right)\right], \label{eq:viscS}
\end{equation}
where $\eta_0$ is a reference viscosity at the reference temperature $T^*$ = 1600~K and $V^*$ is an activation volume. Figure~\ref{fig:visc} shows calculated viscosity profiles for $\eta_0$ = 10$^{21}$~Pa~s and $V^*$ = 2.5, 1.7, and 0.0~cm$^3$~mol$^{-1}$ for the convecting, adiabatic mantle within a 10$M_\oplus$ super-Venus planet, following \citet{Vlada2012}. Our parameterized formulation for stagnant-lid convection is based on numerical modeling with the incompressible fluid approximation using temperature-dependent but pressure-independent viscosity \citep[e.g.,][]{SM2000,JK2009}. Therefore, we are assuming pressure-independent constant potential viscosity, where the effect of temperature increase along an adiabatic gradient exactly balances the effect of pressure on viscosity, which requires a non-zero, positive activation volume. With the linear temperature gradient in Fig.~\ref{fig:visc}, the assumption of constant potential viscosity corresponds to $V^*$~=~0.22~cm$^3$~mol$^{-1}$.

Strongly pressure-dependent viscosity can cause dramatically different behavior to emerge from parameterized convection models, including sluggish lower mantle convection and even the formation of a conductive lid above the core/mantle boundary \citep{Vlada2012}. If convection were effectively suppressed in the lower mantle of large rocky planets, melt production would be significantly decreased and the likelihood of plate tectonics might decrease along with convective vigor as planetary mass increased. The thermal conductivity and expansivity of the mantle are also predicted to increase and decrease, respectively, with depth because of increasing pressure, but the effects of these changes on mantle dynamics are dwarfed by the putative increase in mantle viscosity \citep{Vlada2011,Vlada2012}. 

Although an increase in viscosity under greater pressures seems intuitive, the straightforward application of limited, low-pressure experimental data may not accurately describe the rheology of massive terrestrial planets. In fact, four mechanisms, including a transition from vacancy to interstitial diffusion mechanisms, may cause a viscosity decrease with depth above a pressure of $\sim$0.1 TPa \citep{Karato2011}, as post-perovskite and additional high-pressure mineral phases dominate mantle rheology. According to this study, viscosities in the deep interiors of super-Venus planets may be less than the viscosity of Earth's lower mantle, potentially by as much as 2-3 orders of magnitude. Moreover, the depth-dependence of viscosity within Earth's perovskite-dominated mantle is still debated because Earth's viscosity profile has not been well-constrained.

Despite basic consensus that Earth's lower mantle is probably more viscous than the upper mantle, the magnitude of the viscosity contrast remains controversial. Early studies of Earth's topography and geoid suggested a $\sim$300-fold increase in viscosity between Earth's upper and lower mantle \citep{Hager1989}. On the other hand, analyses of post-glacial rebound predict an order of magnitude less of viscosity increase \citep[e.g.,][]{Kaufmann2002} and gravity data are consistent, albeit loosely, with uniform or only slightly depth-dependent mantle viscosity \citep{Soldati2009}. A joint inversion of these data sets predicts that Earth's internal viscosity increases by $\sim$2 orders of magnitude throughout the mantle \citep{Mitrovica2004}, but it is noted that such inversions are known to suffer from severe nonuniqueness \citep[e.g.,][]{King1995,Kido1997}. In any case, considering the tremendous uncertainty surrounding viscosity profiles within Earth and putative super-Venus planets, it remains legitimate to investigate how large rocky planets might evolve in the limiting case of pressure-independent potential viscosity. 

\subsection{Escaping the Stagnant-Lid Regime}

Terrestrial planet evolution strongly depends on the regime of mantle convection. Assuming that brittle failure limits the strength of the lithosphere, our simulations indicate that the effects of lithosphere hydration dominate the effects of planetary mass on yield and convective stresses. That is, the increase in convective vigor with planetary mass only makes plate tectonics marginally more likely. Modeling results for super-Venus planets, however, suggest two additional mechanisms for escaping the stagnant-lid regime. First, massive terrestrial planets in the stagnant-lid regime feature crustal temperature profiles that enter the stability field of eclogite after crust grows beyond a critical thickness. If a sufficiently large fraction of the total crustal thickness is composed of eclogite, the entire crust could be gravitationally unstable and susceptible to foundering because eclogite is intrinsically denser than mantle peridotite.

On Earth, the phase transition from (metamorphosed) basalt to eclogite primarily occurs in subduction zones. Hydration may thus be important to allowing this phase transition to occur relatively rapidly \citep[e.g.,][]{Ahrens1975}, although this type of kinetic calculation strongly depends on diffusion data that are not well-constrained \citep[e.g.,][]{Namiki1993}. Eclogite is also formed during continent-continent collisions such as the Eurasian and Indian plate collisions \citep{Bucher2002}. Furthermore, the high density of eclogite is theorized to have caused delamitation, foundering, and recycling of relatively thick oceanic lithosphere on Earth during the Archaean \citep{Vlaar1994}. Finally, eclogite may be produced in large mountain ranges through magmatic differentiation \citep{Ducea2002} and pressure-induced phase transition \citep{Sobolev2005} in thick continental crust. Evidence for the strong influence of recent eclogite production and foundering on the topography of the central Andes Mountains has been gathered through geodynamics, petrology, and seismology \citep[e.g.,][]{Kay1996,Beck2002,Sobolev2005,Schurr2006}, as synthesized in a numerical study \citep{Pelletier2010}.  Because massive terrestrial planets have relatively high surface gravity, the phase transition to eclogite will occur at a comparatively shallow depth, making eclogite the stable mineral phase for a large fraction of the crust. The formation of a thick eclogite layer then could cause lithosphere foundering or intermittent plate tectonics, as has been proposed in episodic subduction mechanisms for Venus \citep{Turcotte1993, Fowler1996}. 

Representative temperature profiles for massive terrestrial planets pass through the eclogite stability field for plausible initial conditions. In fact, radiogenic heating and thus crustal temperatures should be greater than calculated with Eq.~\ref{eq:temp_profile}, which would increase the speed of the phase transition to eclogite, because crustal construction mostly occurs early in planetary history. For super-Venus planets with masses greater than $\sim$4$M_\oplus$, eclogite may be the stable phase for the majority of the crust unless the initial mantle potential temperature or magnitude of internal heating is very low. So, crust material may undergo a phase transition to eclogite at relatively shallow depths as the crust grows during thermal evolution in the stagnant-lid regime, forming a thick eclogite layer that could subsequently founder. The buoyant stress from the presence of eclogite scales as $\Delta\rho g h_e$, where $\Delta\rho$ $\sim$ 100 kg m$^{-3}$ is the difference between the densities of eclogite and the mantle. On the other hand, the lithospheric strength scales as $\mu\rho_Lgd_e$, which can only be overcome when the depth scale of the eclogitic layer becomes large enough (at least locally, for example, by foundering). As long as crustal production continues on large rocky planets, eclogite formation and foundering could occur periodically, possibly yielding a regime of mantle convection resembling intermittent plate tectonics. Although we suggest that this process is plausible, pursuing its dynamics in detail is left for future studies.

High surface and crustal temperatures may also cause periodic transitions from the stagnant-lid regime to a form of mobile-lid convection. In this work, massive terrestrial planets in the stagnant-lid regime with surface temperatures held constant at 300 K tend to have very hot crusts. If high surface temperatures exist alongside high crustal temperatures, a transition from stagnant-lid convection to a mobile-lid regime can occur \citep{Reese1999}. Feedback between a changing mantle convection regime and a periodic atmospheric greenhouse effect driven by varying amounts of volcanism, for instance, may be very important to the evolution of Venus \citep{NBS2011}. As surface temperature depends on atmospheric mass and the composition and luminosity of the central star, however, this possibility of escaping the stagnant-lid regime may not be as robust as the first mechanism based on the formation of self-destabilizing crust

\subsection{Limitations of Parameterized Models}

Any parameterized model suffers shortcomings. Steady-state evolution is assumed, for instance, which poorly captures transient events that occur early in planetary evolution such as large impacts \citep{Agnor1999} and the crystallization of a magma ocean \citep[e.g.,][]{Solo93}. Fundamental assumptions such an adiabatic temperature gradient in the mantle and pressure-independent potential viscosity are controversial, and different approaches such as mixing length theory \citep{Wag2011} may be necessary to calculate the thermal structure of planetary interiors if they are not valid. However, our simplified simulations only aim to illuminate the first-order, relative effects of planetary mass on terrestrial planet evolution. Recreating the thermal history of a particular planet would require the introduction of many additional complications. One-dimensional models only return globally averaged values for important quantities, for instance, but mantle plumes, which may upwell from the core/mantle boundary when the core heat flux is positive, are likely important to local magmatism and surface features on terrestrial planets like Mars \citep[e.g.,][]{Weizman2001} and Venus \citep[e.g.,][]{Smrekar2012}. Furthermore, applying a parameterized approach to Venus, where the precise quantity of magmatism is an key output, requires more computationally intensive simulations to benchmark the relevant scaling laws.

\section{Conclusions}

Terrestrial planet evolution is complicated. Although plate tectonics is observed on Earth, the stagnant-lid regime of mantle convection may be most natural for terrestrial planets; at least, it is most common in our Solar System. Thermal evolution models in this study yield first-order, relative conclusions about the evolution of generic terrestrial planets in the stagnant-lid regime. Principal component analysis of simulation results conducted with a wide range of initial conditions captures the relationships between the large number of parameters that describe the interior of a planet. Depending on initial conditions, these planets may have evolved along a variety of paths, featuring different crustal thicknesses and temperatures, interior temperatures, and degrees of mantle processing. To produce specific histories consistent with spacecraft data obtained from Mars and Venus, complications must be added to these simple models.

Properties of massive terrestrial exoplanets are poorly constrained, so questions about the effects of planetary mass on the likelihood of plate tectonics and other important planetary parameters await definitive answers. In this study, we explored what might happen if internal viscosity is not strongly-pressure dependent, the alternative to which has been explored previously using parameterized models. Although convective vigor increases with planetary mass, the likelihood of plate tectonics is only marginally improved. Simple scaling analyses indicate that mantle melt productivity should increase with planetary mass. Because the increase in mantle processing is slow, however, crustal thickness and the relative fraction of processed mantle actually decrease with increasing planetary mass, as thermal evolution simulations confirm. Surface gravity increases with planetary mass, so pressure in the crust of massive terrestrial planets increases relatively rapidly with depth. Plausible temperature profiles favor a phase transition to gravitationally unstable eclogite during normal crustal formation, whereas the basalt to eclogite transformation rarely occurs aside from subduction on Earth. Therefore, thick eclogite layers, along with mobile, hot crustal material, may be important to the evolution of massive terrestrial planets.

\section{Acknowledgments}
CT Space Grant and the George J. Schultz Fellowship from Yale University's Silliman College supported J. O'Rourke. Constructive comments from two anonymous reviewers considerably improved the content and clarity of this manuscript.


\label{lastpage}

\singlespacing

\section{References Cited}


\appendix

\section{Statistical Analysis of Simulation Results}

\setcounter{table}{0}

\begin{table*}[ht]
\begin{center}
\begin{tabular}{c|*{4}{r}|*{4}{r}}
\hline
Parameter & V1 & V2 & Av. & SD & S1 & S2 & Av. & SD \\ \hline
$T_c$ [K] & 0.29 & 0.25 & 1237 & 106 & 0.36 & 0.03 & 1309 & 112 \\
$T_u$ [K] & -0.14 & 0.45 & 1772 & 132 & 0.03 & 0.45 & 1948 & 138 \\
$T_{cm}$ [K] & -0.10 & 0.44 & 3128 & 215 & 0.06 & 0.43 & 3475 & 225 \\
$h_c$ [km] & 0.34 & 0.23 & 115 & 43.1 & 0.34 & 0.22 & 63.0 & 16.6 \\
$h_l$ [km] & -0.36 & 0.20 & 53.6 & 29.1 & -0.34 & 0.17 & 11.1 & 5.83 \\
$h_{ML}$ [km] & -0.39 & 0.20 & 84.1 & 27.7 & -0.37 & 0.21& 20.7 & 6.89 \\
$F_s$ [mW m$^{-2}$] & 0.28 & 0.24 & 50.7 & 14.7 & 0.37 & 0.14 & 204 & 61.6 \\
$F_m$ [mW m$^{-2}$] & 0.29 & 0.01 & 29.6 & 6.58 & 0.37 & 0.07 & 173 & 46.7 \\
log$_{10}(u$) & 0.28 & -0.17 & 0.89 & 0.26 & 0.30 & -0.14 & 2.00 & 0.30 \\
$C^w_{sm}$ & -0.19 & -0.36 & 0.37 & 0.13 & -0.12 & -0.36 & 0.62 & 0.10 \\
$V_{proc}/V_{sm}$ & 0.17 & 0.39 & 1.04 & 0.25 & 0.12 & 0.38 & 0.50 & 0.16 \\
$t_{c,10\%}$ [Gyr] & -0.26 & -0.04 & 0.12 & 0.23 & -0.27 & 0.19 & 0.13 & 0.21 \\
$\log_{10}(\Delta{t_{c,tot}})$ & -0.34 & 0.18 & 8.42 & 0.42 & -0.18 & 0.36 & 8.54 & 0.38 \\
$\lambda_i/\Sigma \lambda_i$ &  0.41 & 0.27 & - & - & 0.42 & 0.30 & - & -  \\  \hline
\end{tabular}
\end{center}
\caption{Principal component basis matrix for Venus (V) and a 10$M_\oplus$ super-Venus planet (S) for the model output after 4.5 Gyr of planetary evolution. Two eigenvectors account for over 65\% of the variance in the normalized and mean subtracted simulation results. The fractions of the cumulative variances for which each principal component accounts, calculated by dividing the principal component eigenvalue by the sum of the eigenvalues for all principal components, are in the bottom row. Output parameters were mean subtracted and normalized using the listed average and standard deviation values.} \label{table:Venus_PCA}
\end{table*}

Parameterized evolution models involve quite a few model parameters. It is important to understand how simulation results depend on a particular choice of model parameters by testing a variety of situations, but at the same time, it becomes difficult to grasp the inflated amount of numerical data. Principal component analysis (PCA) can be used to assess the effective dimensionality of a given data space. Our intention here is to use PCA to extract major features and trends from a large number of simulation results. Each sensitivity analysis consists of $n$ simulations with $m$ output parameters, comprising a data set $D^m_n$. Some parameters, such as $\Delta\eta_w$ and $u$, exhibit orders of magnitudes of variation, and we consider their logarithms because PCA is designed for linear data sets. We normalize the data set as
\begin{equation}
P^m_n=\frac{D^m_n-\mu^m}{\sigma^m},
\end{equation}
where $\mu^m$ is the average value of the \textit{m}-th output parameter,
\begin{equation}
\mu^m = \frac{1}{n}\sum_{i=1}^nD^m_i,
\end{equation}
and $\sigma^m$ is the standard deviation of the \textit{m}-th output parameter,
\begin{equation}
\sigma^m = \left[\frac{1}{n}\sum_{i=1}^n(D^m_i-\mu^m)^2\right]^{1/2}.
\end{equation}

Because the normalized data have zero mean, the covariance matrix $C_P = P^TP$ can be decomposed as $C_P = A^T\cdot\text{diag}[\lambda_1 \ldots \lambda_m]\cdot A$, where $\lambda_i$, the eigenvalues, are ordered so that $\lambda_1\geq\lambda_2\geq\ldots\geq\lambda_m$. The corresponding eigenvectors are the principal components, which account for a progressively decreasing percentage of data variance. Principal components accounting for at least 65\% (an arbitrary threshold) of the total variance are selected for examination to reveal important aspects of simulation results. 

\end{document}